\documentclass[fleqn,usenatbib]{mnras}

\usepackage{newtxtext,newtxmath}

\usepackage[T1]{fontenc}
\usepackage{ae,aecompl}

\usepackage{graphicx}    
\usepackage{amsmath}    
\usepackage{amssymb}    
\usepackage{appendix}

\title[Three dimensional dust mapping of SNRs]{Three dimensional dust mapping of 12 supernovae remnants in the Galactic anticentre}

\author[Bin Yu et al.]{
Bin Yu,$^{1,3}$\thanks{E-mail: yubin2015g@mail.bnu.edu.cn (BY); bchen@ynu.edu.cn (BQC)}
B. Q. Chen,$^{2}$\footnotemark[1]
B. W. Jiang,$^{1}$
and A. Zijlstra$^{3}$
\\
$^{1}$Department of Astronomy, Beijing Normal University, Beijing 100875, P.R.China\\
$^{2}$South-Western Institute for Astronomy Research, Yunnan University, Kunming, Yunnan 650091, P. R. China\\
$^{3}$Jodrell Bank Centre for Astrophysics, School of Physics and Astronomy, University of Manchester, Oxford Road, Manchester M13 9PL, UK\\
}

\date{Accepted XXX. Received YYY; in original form ZZZ}

\pubyear{2017}

\begin{document}
\label{firstpage}
\pagerange{\pageref{firstpage}--\pageref{lastpage}}
\maketitle

\begin{abstract}
We present three dimensional (3D) dust mapping of 12 supernova remnants (SNRs) in the Galactic anti-center  (Galactic longitude $l$ between 150\degr\ and 210\degr) based on a recent 3D interstellar extinction map. The dust distribution of the regions which cover the full extents in the radio continuum for the individual SNRs are discussed. Four SNRs show significant spatial coincidences between molecular clouds (MCs) revealed from the 3D extinction mapping and the corresponding radio features. The results confirm the interactions between these SNRs and their surrounding MCs. Based on these correlations, we provide new distance estimates of the four SNRs, G189.1+3.0 (IC443, $d=1729^{+116}_{-94} \rm \,pc$), G190.9-2.2 ($d=1036^{+17}_{-81} \rm \,pc$), G205.5+0.5 ($d=941^{+96}_{-94}$ or $1257^{+92}_{-101} \rm \,pc$) and G213.0-0.6 ($d=1146^{+79}_{-80} \rm \,pc$). In addition, we find indirect evidences of potential interactions between SNRs and MCs for three other SNRs. New distance constraints are also given for these three SNRs.

\end{abstract}




\section{Introduction}

A large fraction of core-collapse supernova explosions may happen near the molecular clouds (MCs) which gave birth to their progenitor stars \citep{Huang1986ApJ}. For now, among more than 300 Galactic SNRs, about 70 are known or are speculated to have physical contact with their surrounding MCs \citep{Chen2014IAUS}. The surrounding MCs would play a critical role in the evolution of the supernova remnants (SNRs).

Traditionally, OH maser emission has been used to identify the interactions between the SNRs and their surrounding MCs. However, the OH maser emission does not provide specific knowledge about the environment and interaction around the SNR. The detection threshold also restricts its application, since the intensity of this emission may be very weak and we may miss many SNR-MC associations. MCs traced by the CO emissions are also commonly adopted to find spatial correlations with any possible interacting SNRs \citep[e.g.][]{Huang1986ApJ, Jiang2010ApJ,Jeong2012Ap&SS, Chen2014IAUS}. However, in low density regions of a MC, CO could be undetectable due to the low column density. Besides, there is a substantial fraction of ``CO dark gas'' where $\rm H_2$ cannot be traced by any CO emission \citep{Chen2015MNRAS.448.2187C, Planck2011A&A...536A..19P}. In such cases, dust observation via optical and near-infrared (IR)  extinction seems to be a better way to trace the MCs \citep{Goodman2009ApJ...692...91G,Chen2014MNRAS}. From the three-dimensional (3D) dust extinction mapping, \citet{Chen2017MNRAS.472.3924C} have identified a new dust feature, namely the ``S147 dust cloud'', that may be possibly interacting with the SNR S147. In addition, they have also obtained a new accurate distance estimate of S147 from the 3D extinction analysis.

The distance is key to understanding the properties, such as the age and size, of SNRs. Many methods have been applied to estimate the distances of SNRs. The empirical relation between surface brightness ($\Sigma$) and diameter ($D$) of SNRs, $\Sigma=aD^b$, is often adopted to obtain distances of SNRs. The method is straightforward and can be easily applied. However, the method suffers from large uncertainties, of up to 40\,percent \citep{Case1998ApJ...504..761C}. The empirical relation can be influenced by interstellar medium. The distribution of explosion energy and determination of diameter also cause uncertainties \citep{Guseinov2003A&AT,Zhu2014IAUS}. Another method widely used is to determine the kinematic distance of an SNR based on the HI and CO absorption lines. However, the resulted kinematic distance, especially that in the direction of the Galactic anticentre, also suffers from large uncertainties as the method relies on the Milky Way rotation curve model and the deviations from noncircular rotation \citep{Leahy2008AJ,Tian2007A&A,Tian2008ApJ,Tian2012MNRAS}. Other methods, such as calculating the Sedov distance of shell type SNRs with X-ray observation and association with objects with known distance are restricted to specific SNRs. Recently, estimating the distances of SNRs from the 3D dust mapping has received increasing attention. \citet{Zhao2018ApJ} have obtained the distance of the Monoceros SNR using a 3D dust extinction analysis based on data from the 2MASS photometric survey and the APOGEE and LAMOST spectroscopic surveys. \citet{Shan2018ApJS} have measured distances of 15 SNRs based on the extinction profiles obtained from 2MASS red clumps stars.

In this work, we conduct a systematic study of 12 SNRs in the Galactic anti-center based on the recent 3D extinction maps from  \citet{chen2019MNRAS.483.4277C}. We present a detailed analysis of 3D dust distributions in regions of these SNRs and seeking for MCs which would have spatial correlation with their radio features. It may help us to identify interactions between SNRs and MCs and provide new distance estimates of the corresponding SNRs.

This paper is organized as follows. In Section\,2 and 3 we describe our data and method. Results are presented in Section\,4, and implications, comparisons with other results and further development are discussed and summarized in Section\,5.

\section{Data}
\label{sec:dm}
\citet{chen2019MNRAS.483.4277C} build a new 3D interstellar dust extinction map of the Galactic plane (Galactic longitude $0^{\circ}< l < 360^{\circ}$ and latitude $|b| < 10^{\circ}$). The spatial angular resolution is 6 arcmin. The maps are based on the robust parallax estimates and the high-quality optical photometry from the Gaia Data Release 2 (Gaia DR2), together with the infrared photometry from the 2MASS and WISE surveys.
Color excesses, $E(G-K_{\rm S}), E(G_{\rm BP}-G_{\rm RP}), E(H-K_{\rm S})$ of over 56 million stars are estimated from Random Forest regressions. The distances to individual stars are adopted from \citet{Bailer-Jones2018AJ....156...58B}, who use a Bayesian procedure to transfer the Gaia parallax into distances. In this work, only the distances of stars with Gaia parallax errors no more than 20 percent are adopted. We select individual stars from \citet{chen2019MNRAS.483.4277C} inside 12 squares which centered at our targets SNRs. The side length of each square is decided from the size of the SNR summarized by \citet{GreenCatalog}. The coverages and numbers of sources are presented in Table \ref{tab: coverage}.

\begin{table}
    \centering
    \caption{The coverages and numbers of sources in target SNRs.}
    \label{tab: coverage}
    \begin{tabular}{ccccc} 
        \hline
        ID & l & b & Side Length(arcdeg) &  Number of Sources \\
        \hline
        1 & 189.1 & 3 & 3.0 & 27021\\
        2 & 190.9 & -2.2 & 4.0 & 43092\\
        3 & 205.5 & 0.5 & 4.4 & 62762\\
        4 & 213 & -0.6 & 4.0 & 60939\\
        5 & 182.4 & 4.3 & 1.2 & 4227\\
        6 & 152.4 & -2.1 & 3.0 & 27130\\
        7 & 160.9 & 2.6 & 4.0 & 50815\\
        8 & 206.9 & 2.3 & 1.6 & 9431\\
        9 & 156.2 & 5.7 & 3.0 & 21518\\
        10 & 166 & 4.3 & 2.0 & 13664\\
        11 & 178.2 & -4.2 & 1.6 & 6549\\
        12 & 179 & 2.6 & 1.2 & 5367\\

        \hline
    \end{tabular}
\end{table}

To demonstrate the spatial coincidence between the MCs and radio features of SNRs, radio data from the Sino-German 6\,cm Polarization Survey of the Galactic Plane \citep{Gao2010A&A...515A..64G} are adopted. This survey was performed using a dual-channel 6\,cm receiving system on the Urumqi 25 m radio telescope. The receiving system has a system temperature of about 22 K, centers at 4800 MHz. The half power beam width (HPBW) is $9.5'$.

\section{Method}

To identify spatial correlations between SNRs and MCs, we first analyze the dust spatial distribution in three dimensions for regions of the individual SNRs. The method is similar to that of \citet{chen2019MNRAS.483.4277C}. For each SNR, we divide the sample of stars into pixels of size $9^{\prime} \times 9^{\prime}$. For each pixel, the dust extinction profile is parameterised by a piecewise linear function,
\begin{equation}
E(\mu)=\sum_{i=0}^{\mu}(\Delta E^i),
\label{eq:ar0}
\end{equation}
\noindent where $\mu$ is the distance module, $\Delta E^i$ is color excess in the $i$th distance bin and $E$ refers to
$E(G-K_{\rm S})$, which is more sensitive to the amount of the interstellar dust. The length of each bin is
set as $\delta\mu$  = 0.5\,mag.  A Markov chain Monte Carlo (MCMC) procedure is performed to derive the best set of
$\Delta E^i$ which have the maximum value of the likelihood defined as,
\begin{equation}
L = \Pi_{n=1}^{N} \frac{1}{\sqrt{2\pi} \sigma_n}{\exp}(\frac{-(E^n_{\rm obs}-E^n_{\rm mod})^2}{2\sigma^2_n}),
\end{equation}
where $n$ is index of star in the pixel,
$E^n_{\rm obs}$ and $E^n_{\rm mod}$ are respectively the colour excess from \citet{chen2019MNRAS.483.4277C}
and that given by Eq.~(1) of the star, $\sigma_n$ is the combined
uncertainty of the derived colour excess $E^n_{\rm obs}$ and distance $d^n$,
given by $\sigma_n = \sqrt{\sigma_{E^n_{\rm obs}}^2+(E^n_{\rm obs}\dfrac{\sigma_{d^n}}{d^n})^2}$,
and $N$ is the total number of stars in the pixel.
With the resulted $\Delta E^i$, we then plot the spatial distribution of the differential dust extinction for the individual
dust slices of each SNR and compare to the morphology of the 6\,cm radio emissions of the SNR to identify any MCs
that have possible spatial correlation with the SNR.

Once we find any possible correlations between the morphologies of MCs and SNRs, a method similar as that from
\citet{Chen2017MNRAS.472.3924C} is then adopted to obtain the accurate distance of the MC. We select all stars in the region
of the MC and examine the variation of their colour excess values with distances. As the dust density in the MC is higher than
that in the diffuse medium, the colour excess will increase sharply when meeting with MC.
 Thus the distance to the MC can then be determined from the position where the extinction increases sharply.
 A simple colour excess model $E(d)$ is adopted to find the position, by
 \begin{equation}
E(d)=E^{0}(d)+E^{1}(d),
\end{equation}
where $E^{0}(d)$ is the colour excess from the diffuse medium and $E^1(d)$ is that contributed from the MC.
Similar to in \citet{Chen2017MNRAS.472.3924C}, the models of $E^0(d)$ and $E^1(d)$ are respectively defined by,
\begin{equation}
E^{0}(d)=ad+bd^2,
\label{eq:ar0}
\end{equation}
and
\begin{equation}
    E^{1}(d)=\frac{\delta E}{2}\left(1+{\rm erf}\left(\frac{d-d_0}{\sqrt{2}D\cdot d_0}\right)\right),
    \label{eq:ar1}
\end{equation}
where $a$ and $b$ are polynomial coefficients, $\delta E$ is the increasing amplitude of colour excess, $D$
is the width of the corresponding SNR and $d_0$ is the position where the sharp colour excess increasing occurs (i.e. the distance of the MC).
Since the distance uncertainties could be larger than the width of the MC, it is very hard to constrain the width of the MC.
Thus we simply assume that the width of the MC equals the width of the correlated SNR.
For each MC, we bin the stars in distance intervals with width of 0.1\,kpc and fit the medium colour
excess values in each bin with the model described above using a Monte Carlo procedure.

\section{Results}
\label{sec:results}
3D colour excess distribution maps are plotted as contours at several 0.5\,kpc distance intervals between 0 and 4\,kpc in the directions of different SNRs. The whole maps are available online. We select the most distinguishable MCs and compare to the 6\,cm radio emission features of SNRs to find any spatial correlations. If the morphology of a MC is spatially coincident or anti-coincident with the radio observations, we assume that there would be possible interaction between the MC and the corresponding SNR. Distance of the SNR is then given by fitting the colour excess profile of the stars in the MC area. Results of the individual SNRs are described in detail below.

\subsection{G189.1+3.0, IC443}
IC443 (G189.1+3.0) is located near the Gem OB1 association \citep{Cornett1977A&A....54..889C}. A pulsar associated with IC443 \citep{Olbert2001ApJ...554L.205O} and a nearby dense giant molecular cloud with OH maser emission \citep{Claussen1997ApJ...489..143C} strongly suggest a core-collapse origin. However, \citet{Leahy2004AJ....127.2277L} argue that there is a separate supernova remnant, G189.6+3.3, and the pulsar is more likely related to it rather than to IC 443. There are plenty of evidences for the interaction between IC443 and its nearby MC. \citet{Denoyer1978MNRAS.183..187D} first detect high-velocity HI emission and broad emission in the region. The absorption lines of CO and OH are then detected by \citet{Denoyer1978MNRAS.183..187D, Denoyer1979ApJ...228L..41D, Denoyer1979ApJ...232L.165D}. Shocked HI filaments \citep{Braun1986A&A...164..193B, Lee2008AJ....135..796L} and shocked molecular clumps \citep{Huang1986ApJ...302L..63H, Dickman1992ApJ...400..203D, Snell2005ApJ...620..758S} have also been found in the SNR. The OH (1720\,MHz) masers coincident with the shocked molecular material, observed by \citet{Claussen1997ApJ...489..143C} and \citet{Hewitt2006ApJ...652.1288H} also indicate the SNR-MC interaction. The distance of IC443 is estimated as 0.7-1.5\,kpc, suggested by mean optical velocity, or 1.5-2\,kpc, given by the association with the $\rm H{\sc II}$ region S249 \citep{GreenCatalog}.

\begin{figure*}
    \includegraphics[width=17.5cm]{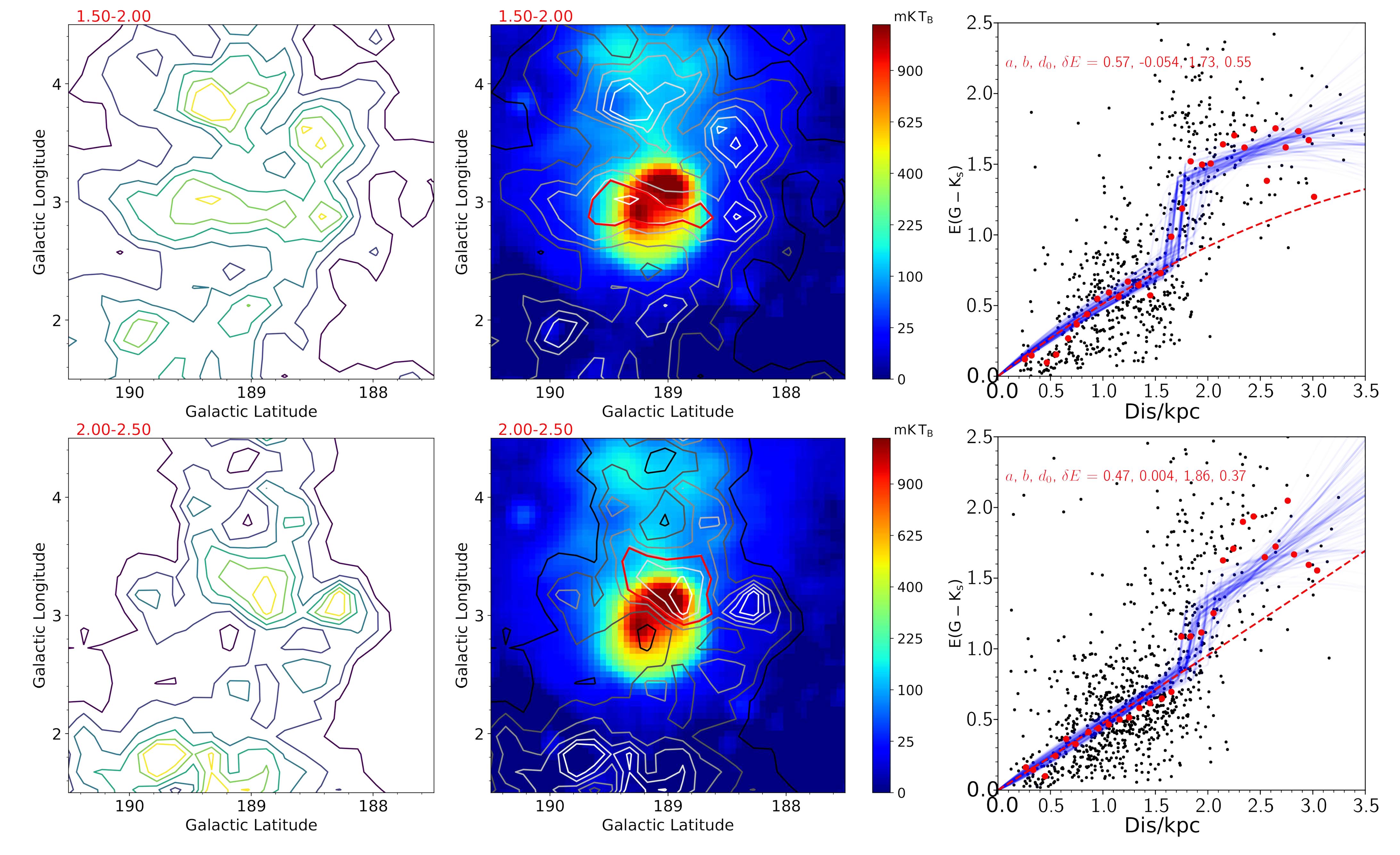}
    \caption{{\it Left:} differential colour excess distribution maps of IC443 at distance bin 1.5-2.0\,kpc (upper panel) and 2.0-2.5\,kpc (bottom panel). {\it Middle: } the comparisons of the differential colour excess distribution maps of IC443 at distance bin 1.5-2.0\,kpc (upper panel) and 2.0-2.5\,kpc (bottom panel) with the morphologies of the 6\,cm radio emissions. {\it Right:} the colour excess profile fitting for two MCs (MC1a, top panel and MC1b, bottom panel). The colour contours in the left panels and the grey-scale contours in the middle panels represent the distribution of the differential colour excesses, which vary from 0.2\,mag to 0.7\,mag with interval of 0.1\,mag. The colour maps in the middle panels represent the Urumqi 6\,cm radio image of IC443. The red polygons in the middle panels represent the corresponding regions of MC1a and MC1b we selected. Red points are median values in each distance bin. Red dashed lines are extinction curve without the contribution of MCs. Blue lines are the best-fit colour excess profiles resulted from 100 randomly picked sample by the Monte Carlo analysis.}
    \label{fig:169_a}
\end{figure*}

In the region of IC443, significant MC features are found at distance bins 1.5-2.0\,kpc and 2.0-2.5\,kpc (first column in Fig.~\ref{fig:169_a}). The MC features have a significant spatial coincidence with the morphology of the 6\,cm radio observation of IC443. One feature is on the left-bottom side ($l \sim 189.4 ^{\circ}, b \sim 3.0 ^{\circ}$, noted as MC1a), and another in the right-top corner ($l \sim 189.0^{\circ}$, $b \sim 3.2^{\circ}$, noted MC1b). Our result is consistent with the previous works mentioned above, which confirms that these MCs are interacting with IC443. We select stars in regions of the two MC features, which are indicated by the red polygon in the second columns of Fig.~\ref{fig:169_a} and then calculate the distances of the two MCs. Results are shown in the third column of Fig.~\ref{fig:169_a}. Our results yield $d_0=1729^{+116}_{-94} \rm\,pc$, $\delta E(G-K_{\rm S})=0.5_{-0.13}^{+0.10}\rm \,mag$ for the MC1a, and $d_0=1865^{+174}_{-116}\rm \,pc$, $\delta E(G-K_{\rm S})=0.37_{-0.15}^{+0.16}\rm \,mag$ for MC1b. The distances of the two MCs are consistent with each other. From the colour excess profile fitting diagram of MC1b plotted in the bottom-right panel of Fig.~\ref{fig:169_a}, we are able to identify two significant sharp increases of the colour excess, one at about 1.7\,kpc and the other at about 2.0\,kpc. Considering that there is also a MC feature in the direction of S264 ($l \sim 189.0^{\circ}, b \sim 4.2^{\circ}$) at distance bin 1.5-2.0\,kpc, which is suggested to correlate with IC443, we suggest that the two clouds at distance between 1.5-2.0\,kpc are more likely to be correlated with IC443. The result, $d_0=1729^{+116}_{-94}\rm \,pc$, is adopted as the distance of IC443 in our work.

\subsection{G190.9-2.2}
\citet{Foster2013A&A...549A.107F} identify G190.9-2.2 as a shell-type remnant. Its east and west halves are elongated perpendicular to the Galactic plane. They report that the SNR is evolving in a low-density region bounded by two MCs traced by $\rm ^{12}CO$ in the east and west of the SNR. The surface brightness of this SNR is very low ($\rm \sum_{1GHz}\le5\times10^{-23}\, W\,m^{-2}\,Hz^{-1}\,sr^{-1}$). According to the kinematic distance to the remnant of $\rm 1.0\pm0.3\,kpc$ based on the CO and HI observations (radial velocity of $\rm +5.1\,km\,s^{-1}$), the physical dimension is about $\rm 18\times16\,pc$.

\begin{figure*}
    \includegraphics[width=17.5cm]{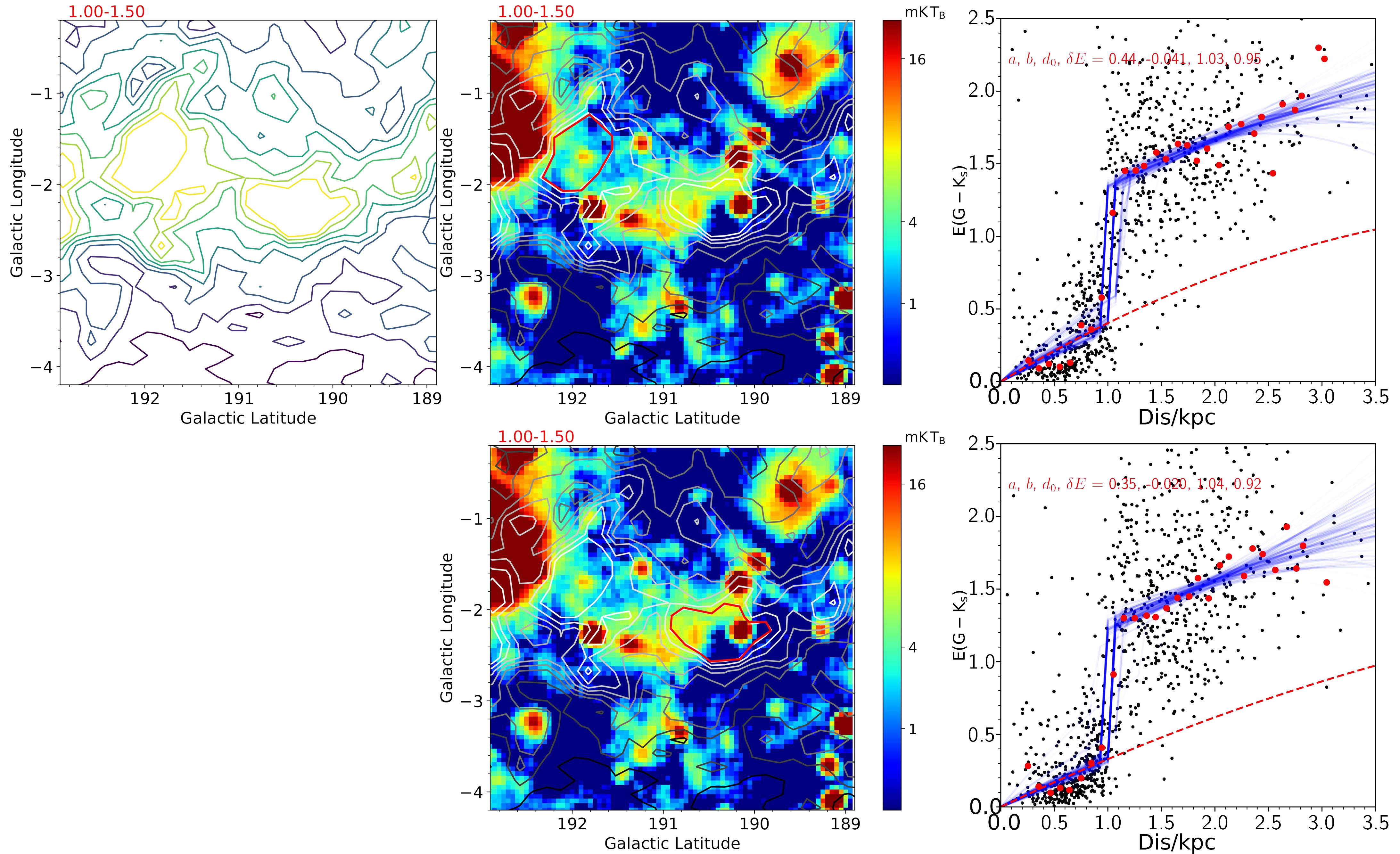}
    \caption{{\it Left:} the colour excess distribution map of G190.9-2.2 at distance bin 1.0-1.5\,kpc . {\it Middle: } two high reddening regions selected from the map, MC2a(upper panel) and Mc2b (bottom panel) with the morphologies of the 6\,cm radio emissions. {\it Right:} the colour excess profile fitting for two MCs (MC2a, top panel and MC2b, bottom panel).  The colour contours in the left panels and the grey-scale contours in the middle panels represent the distribution of the differential colour excesses, which vary from 0.18\,mag to 0.79\,mag with interval of 0.09\,mag. The colour maps in the middle panels represent the Urumqi 6\,cm radio image of G190.9-2.2. The red polygons in the middle panels represent the corresponding regions of MC2a and MC2b we selected. Red points are median values in each distance bin. Red dashed lines are extinction curve without the contribution of MCs. Blue lines are the best-fit colour excess profiles resulted from 100 randomly picked sample by the Monte Carlo analysis.}
    \label{fig:170_a}
\end{figure*}

In the 3D extinction maps of G190.9-2.2, we find two MC features, which are located respectively on the east and west side of the remnant ($l \sim 192.0^{\circ}$, $b \sim -1.6^{\circ}$, noted MC2a, $l \sim 190.4^{\circ}$, $b \sim -2.2^{\circ}$, noted MC2b, left panel of Fig.~\ref{fig:170_a}) and are in good agreement of the structures suggested by \citet{Foster2013A&A...549A.107F}. Thus we suggest the two MCs may be interacting with the SNR. These two clouds are located at distance bin 1.0-1.5\,kpc. Stars in the regions of the two MCs are then selected and distances of the individual MCs are fitted separately.  We obtain $d_0=1031^{+19}_{-79}\rm \,pc$, $\delta E(G-K_{\rm S})=0.95_{-0.11}^{+0.05}\rm \,mag$ for MC2a and $d_0=1036^{+17}_{-81}\rm \,pc$ and $\delta E(G-K_{\rm S})=0.92_{-0.08}^{+0.05}\rm \,mag$ for MC2b. The distances of the two MCs are consistent with each other. Thus the distance of the SNR is estimated as the averaged distance of the two MCs, which is $1.03^{+0.02}_{-0.08}\rm \,kpc$.

\subsection{G205.5+0.5, Monoceros Nebula}
The Monoceros SNR is a large shell-type SNR. The bright Rosette Nebula is located in the south edge of the SNR shell and in the north there is an H{\sc II} region, Sh 2-273 \citep{Graham1982A&A...109..145G}, which has an average spectral index $\alpha=-0.47\pm 0.06$ ($S_\nu \sim \nu^\alpha$) from 111-2700\,MHz radio data. \citet{Leahy1986A&A...156..191L} suggest that the SNR is expanding in a low-density (0.003\,$\rm cm^{-3}$) medium. There is large discrepancy of distance estimates of the Monoceros Nebula between different studies. For example, the mean optical velocity of the SNR suggests a distance of 0.8\,kpc, but the low-frequency radio absorption gives an estimate of 1.6\,kpc \citep{GreenCatalog}. Based on all the available information, \citet{Odegard1986ApJ...301..813O} argue that the Monoceros Nebula is located at the same distance of the Monoceros OB2 association, which is about 1.6\,kpc, and they obtain a diameter of about 106\,pc for the SNR. Interactions between the Monoceros and the Rosette Nebulae have been identified at different wavelengths \citep{Fountain1979ApJ...229..971F, Jaffe1997ApJ...484L.129J, Fiasson2008ICRC....2..719F, Torres2003PhR...382..303T,Xiao2012A&A...545A..86X}. \citet{Zhao2018ApJ} obtain distances of the Monoceros SNR and the Rosette Nebula of 1.98\,kpc and 1.55\,kpc, respectively, using a 3D extinction analysis based on data from the 2MASS photometric survey and the APOGEE and LAMOST spectroscopic surveys. They argue that there is no interaction between the two nebulae.

\begin{figure*}
    \includegraphics[width=17.5cm]{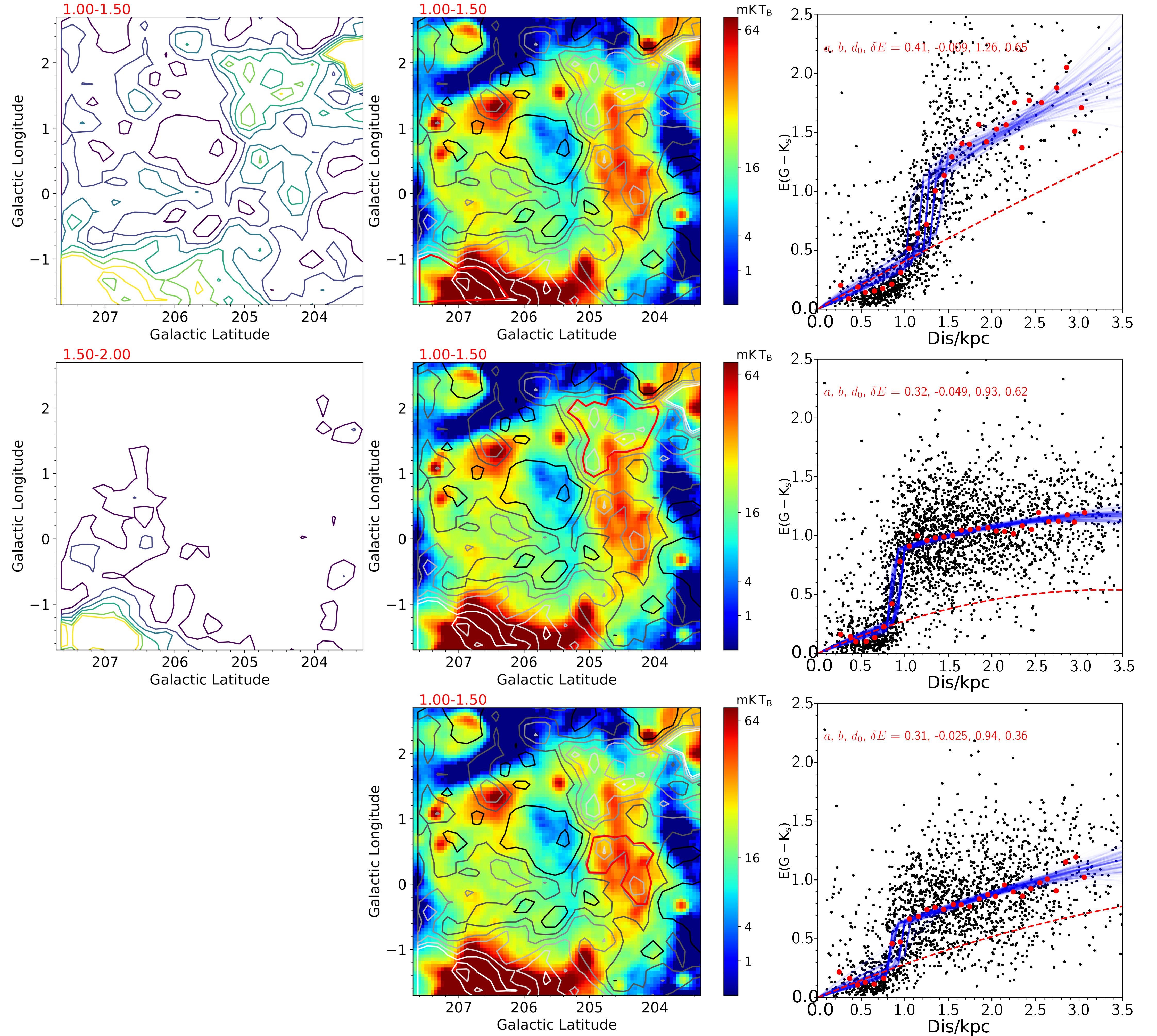}
    \caption{{\it Left:} differential colour excess distribution maps of the Monoceros Nebula at distance bin 1.0-1.5\,kpc (upper panel) and 1.5-2.0\,kpc (bottom panel). {\it Middle: } three high reddening regions selected from the map, MC3a (upper panel), MC3b (middle panel) and MC3c (bottom panel), with the morphologies of the 6\,cm radio emissions. The colour contours in the left panels and the grey-scale contours in the middle panels represent the distribution of the differential colour excesses, which vary from 0.18\,mag to 0.63\,mag with interval of 0.09\,mag. The colour maps in the middle panels represent the Urumqi 6\,cm radio image of the Monoceros Nebula. The red polygons in the middle panels represent the corresponding regions of MC3a, MC3b and MC3c we selected. Red points are median values in each distance bin. Red dashed lines are extinction curve without the contribution of MCs. Blue lines are the best-fit colour excess profiles resulted from 100 randomly picked sample by the Monte Carlo analysis.}
    \label{fig:172_a}
\end{figure*}

In this work, we find that the MC feature at distance between 1.0 and 1.5\,kpc in the direction of the SNR is coherent surrounding the remnant (the top left panel of Fig.~\ref{fig:172_a}). MC features are visible at both the directions of the Rosette Nebula and the H{\sc II} region Sh 2-273. Furthermore, there are MC features of relatively lower colour excesses located at the eastern and western shell of the Monoceros Nebula. We select three MCs, two in the west ($l \sim 204.8^{\circ}$, $b \sim 1.6^{\circ}$, noted MC3a, $l \sim 204.6^{\circ}$, $b \sim 0.4^{\circ}$, noted MC3b) and the other in the direction of the Rosette Nebula (MC3c). Both the distances of MC3a and MC3b are suggested as $\sim$ 0.9\,kpc in the colour excess profile fitting procedure. The best-fit results of MC3a are $d_0=927^{+12}_{-75} \rm\,pc$ and $\delta E(G-K_{\rm S})=0.62_{-0.04}^{+0.04} \rm\,mag$ and those of MC3b are $d_0=941^{+96}_{-94} \rm\,pc$ and $\delta E(G-K_{\rm S})=0.36_{-0.09}^{+0.05} \rm\,mag$. However, the colour excess profile in regions of MC3c, which correlates to the Rosette Nebula is not consistent with MC3a and MC3b. The resulted distance, $d_0$ = 1.26\,kpc, is larger than that of MC3a and MC3b. However, this agrees with the previously determined distance of the Rosette Nebula by taking the uncertainties into account. For example,  e.g. \citet{Zhao2018ApJ} obtained 1.55\,kpc, and \citet{Davies1978A&AS...31..271D} obtained 1.6\,kpc for the Rosette Nebula. We consider that MC3c is the Rosette Nebula and different from the MC3a and MC3b MCs. Therefore, if the Monoceros Nebula is interacting with the MC3a and MC3b MCs, its distance can be the average of MC3a and MC3b, i.e. $0.93^{+0.05}_{-0.08} \rm\,kpc$. On the other hand, if it is interacting with the Rosette Nebula, its distance is then $1.26^{+0.09}_{-0.10} \rm\,kpc$. As described in the above paragraph, some works identified the interaction between the Monoceros and Rosette Nebulae, it is also possible that the distance of Monoceros SNR to be 1.26\,kpc, and MC3a and MC3b are foreground nebulae. Nevertheless, there is some discrepancy with the result (1.98\,kpc) of \citet{Zhao2018ApJ} even when the uncertainties of both results are considered. The main reason lies in that we take the distance of the surrounding MC as the distance of the SNR while \citet{Zhao2018ApJ} derive the distance from the stars within the SNR region, and their study shows that the Monoceros SNR is about 0.4\,kpc more distant than the Rosette Nebula. Which one is correct depends on if the SNR is really interacting with the MC.


\subsection{G213.0-0.6}
G213.0-0.6 is located to the east of H{\sc II} region S248. It is an old SNR, with a partially shell-like structure and extremely low radio surface brightness \citep{Reich2003A&A...408..961R}. The structure of the 6\,cm radio emission in this SNR is very fragmented. \citet{Su2017ApJ...836..211S} combine the morphological correspondence between CO and radio observations and the broadening of CO profiles to conclude that molecular gas at radial velocity $V_{\rm LSR}$ $\sim 9 \rm \,km\,s^{-1}$ would be physically associated with the remnant. They obtain a kinematic distance to the remnant of about 1.0\,kpc, which is consistent with the estimates from the updated $\sum-D$ relationship \citep{Pavlovic2014SerAJ.189...25P} and the 3D extinction map \citep{Green2015ApJ...810...25G}. They also argue that the H{\sc II} region Sh2-284, which is located near the southwestern border of the SNR, has a distance of 4-5\,kpc. Therefore there is no correlation between Sh2-284 and the remnant.

 We find some segmented MC features that may correlate to the morphology of 6\,cm radio observation of the SNR at distance bin between 1.0 and 1.5\,kpc. From the colour excess profile fitting, we obtain $d_0$ = $1146^{+79}_{-80}$\,pc and $\delta E(G-K_{\rm S})=0.35_{-0.03}^{+0.02} \rm\,mag$. Our estimate of distance is consistent with that from \citet{Su2017ApJ...836..211S}.

\begin{figure*}
    \includegraphics[width=17.5cm]{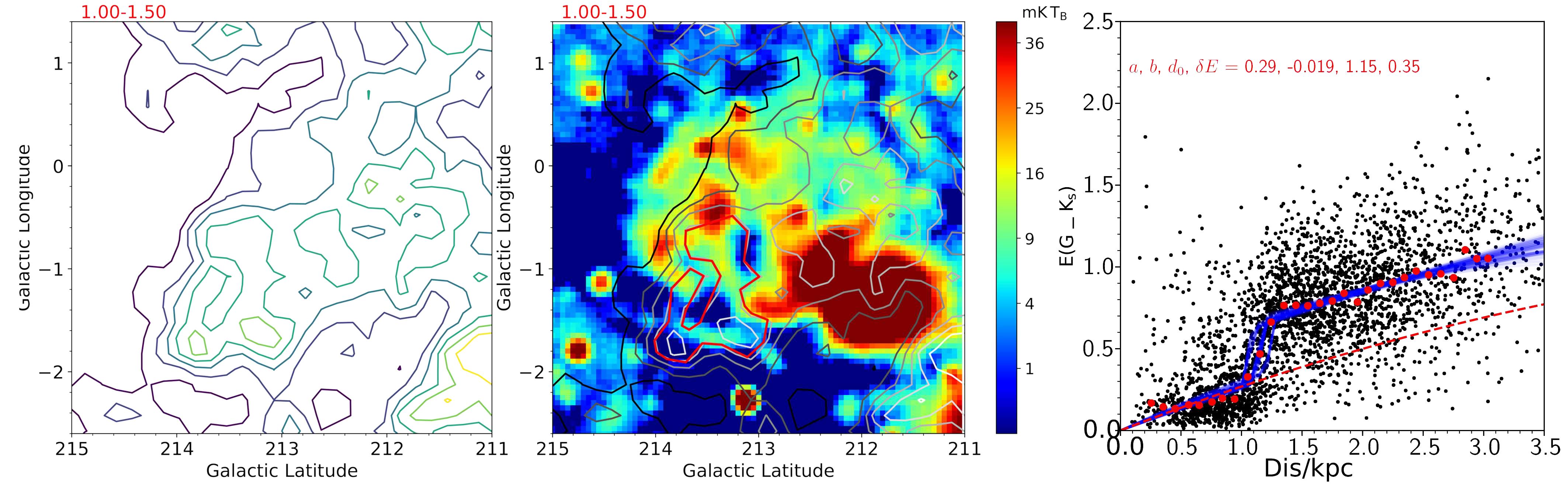}
    \caption{{\it Left:} the colour excess distribution map of G213.0-0.6 at distance bin 1.0-1.5\,kpc . {\it Middle: } the reddening region selected from the map, with the morphologies of the 6\,cm radio emissions. {\it Right:} the colour excess profile fitting for the MC. The colour contours in the left panel and the grey-scale contours in the middle panel represent the distribution of the differential colour excesses, which vary from 0.16\,mag to 0.57\,mag with interval of 0.08\,mag. The colour map in the middle panel represents the Urumqi 6\,cm radio image of G213.0-0.6. The red polygon in the middle panel represent the corresponding region of MC we selected. Red points are median values in each distance bin. Red dashed lines are extinction curve without the contribution of MCs. Blue lines are the best-fit colour excess profiles resulted from 100 randomly picked sample by the Monte Carlo analysis.}
    \label{fig:174_a}
\end{figure*}

\subsection{G182.4+4.3}
G182.4+4.3 was first detected at radio wavelength by \citet{Kothes1998A&A...331..661K}. They report that the remnant has a shell structure in the southwest. The southern shell is bright and has a circular shape. In contrast, the northern shell is much fainter and flattened. From a discussion of their radio observations at 1400\,MHz, 2675\,MHz, 4850\,MHz, and at 10450\,MHz, they argue that G182.4+4.3 is at a distance of $\geq 3$\,kpc, having a radius of about 22.5\,pc, and probably expanding according to the classical Sedov equations. They have checked the ROSAT All-Sky Survey and found no visible X-ray emission detected in the region of the remnant.

\begin{figure*}
    \includegraphics[width=17.5cm]{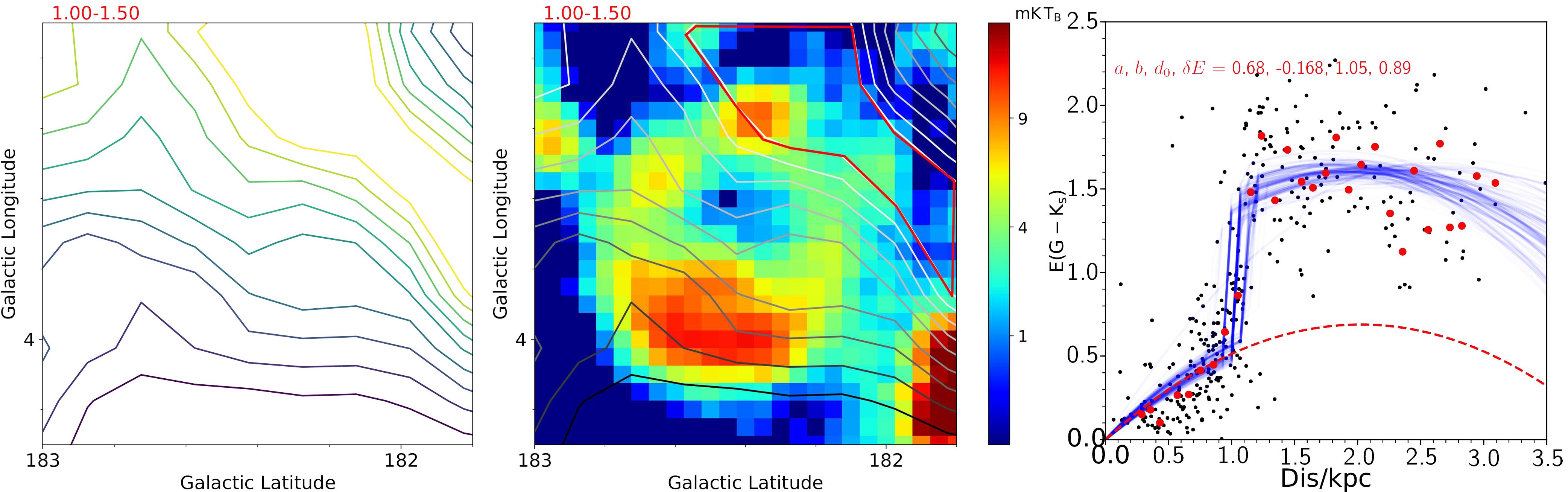}
    \caption{{\it Left:} the colour excess distribution map of G182.4+4.3 at distance bin 1.0-1.5\,kpc. {\it Middle: } the reddening region selected from the map, with the morphologies of the 6\,cm radio emissions. {\it Right:} the colour excess profile fitting for the MC. The colour contours in the left panel and the grey-scale contours in the middle panel represent the distribution of the differential colour excesses, which vary from 0.14\,mag to 0.70\,mag with interval of 0.07\,mag. The colour map in the middle panel represents the Urumqi 6\,cm radio image of G182.4+4.3. The red polygon in the middle panel represent the corresponding region of MC we selected. Red points are median values in each distance bin. Red dashed lines are extinction curve without the contribution of MCs. Blue lines are the best-fit colour excess profiles resulted from 100 randomly picked sample by the Monte Carlo analysis.}
    \label{fig:167_a}
\end{figure*}

In this work, we only find one MC at the distance bin 1.0-1.5\,kpc (top left panel in Fig.~\ref{fig:167_a}). Beyond 1.5\,kpc, there are barely any cloud features. The MC we find at the north and northwest sides of the SNR has no good spatial correlation with the remnant. However, an arc-shaped cloud has also been detected by the CO survey conducted by \citet{Jeong2012Ap&SS} in the same direction. According to the well-matched borders of the MC and the faint northern shell of the SNR, \citet{Jeong2012Ap&SS} suggest that the MC would be blocking the expansion of remnant. We have calculated the distance of the cloud, which yields a distance of $d_0$ = $1050^{+87}_{-91}$\,pc and $\delta E(G-K_{\rm S})=0.89_{-0.11}^{+0.08} \rm\,mag$. If the blocking scenario is true, the SNR may be located at the same distance.

\subsection{G152.4-2.1}
\begin{figure*}
    \includegraphics[width=17.5cm]{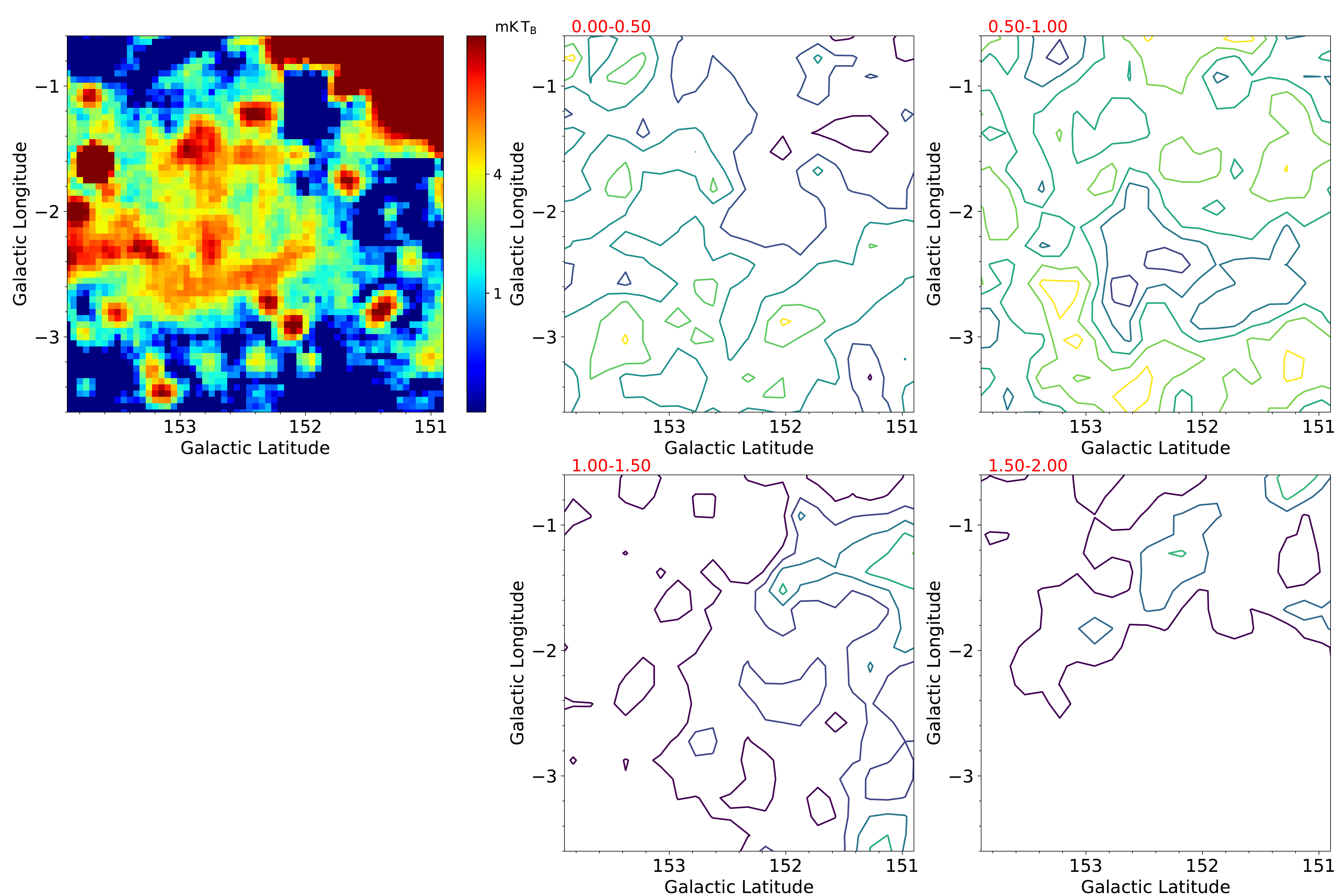}
    \caption{{\it Panel 1:} the morphologies of the 6\,cm radio emissions of G152.4-2.1. {\it Panel 2-5:} the colour excess distribution maps of G152.4-2.1 at distance bin 0.0-0.5\,kpc, 0.5-1.0\,kpc, 1.0-1.5\,kpc, 1.5-2.0\,kpc. The colour contours in the middle and right panels represent the distribution of the differential colour excesses, which vary from 0.22\,mag to 0.78\,mag with interval of 0.11\,mag. The colour map in the left panel represents the Urumqi 6\,cm radio image of G152.4-2.1.}
    \label{fig:159_a}
\end{figure*}

G152.4-2.1 was first discovered and identified as a SNR by \citet{Foster2013A&A...549A.107F} from radio observations. It has an integrated radio continuum spectral index of $-0.65\pm0.05$. There are two radio-bright shells in the north and south, paralleling the Galactic plane. The very bright H{\sc II} region Sh2-206 is seen in the northwest corner (as shown in the first panel of Fig.~\ref{fig:159_a}). The distance of the SNR is estimated to be about 1.1\,kpc with a large uncertainty.

In this work, we find several MCs at nearby distances $d$ $\leq$ 1.0\,kpc in the direction of the SNR. However, none of the MCs has good spatial correlation with the morphology of the radio observation of the SNR. There is no significant evidence for the interaction between MCs and the SNR. Beyond 1.0\,kpc, there is little evidence for MCs. If there are any MCs that are interacting with the remnant, the distance should be $\leq$ 1.0\,kpc.

\subsection{G160.9+2.6, HB9}
HB9 is a bright large nearby SNR. An arc-shaped shell is shown in the east part of the remnant. Inside the remnant, filamentary structures are visible. From the 6\,cm radio emission map, the shells are clearly visible. A spectral index of $\alpha~=~-0.57\pm0.03$ is presented based on a TT-plot between 865\,MHz data and 4750\,MHz Effelsberg data \citep{Reich2003A&A...408..961R}. \citet{Leahy2007A&A...461.1013L} estimate a kinematic distance of the SNR as $d$ = $0.8\pm0.4$\,kpc. Instead, \citet{Leahy1991AJ....101.1033L} argue that HB9 should be located near the MCs Sh217 and Sh219, but there is no definitive evidence for any interactions between the SNR and any MCs for now.

\begin{figure*}
    \includegraphics[width=17.5cm]{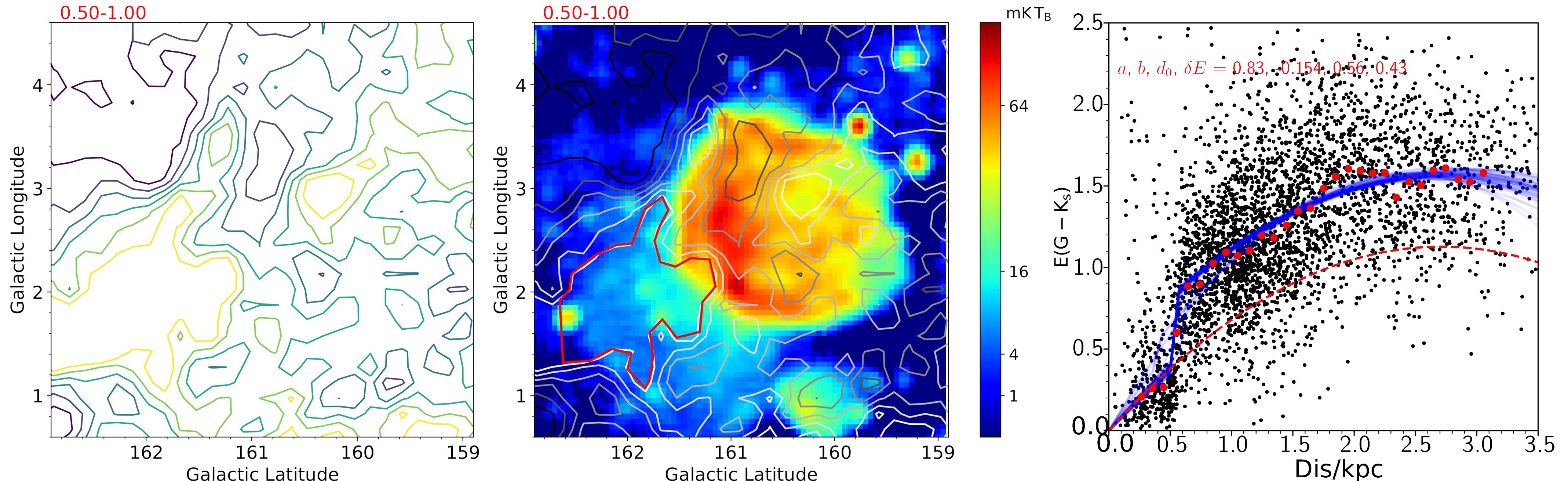}
    \caption{{\it Left:} the colour excess distribution map of HB9 at distance bin 0.5-1.0\,kpc. {\it Middle: } the reddening region selected from the map, with the morphologies of the 6\,cm radio emissions. {\it Right:} the colour excess profile fitting for the MC. The colour contours in the left panel and the grey-scale contours in the middle panel represent the distribution of the differential colour excesses, which vary from 0.17\,mag to 0.60\,mag with interval of 0.09\,mag. The colour map in the middle panel represents the Urumqi 6\,cm radio image of HB9. The red polygon in the middle panel represent the corresponding region of MC we selected. Red points are median values in each distance bin. Red dashed lines are extinction curve without the contribution of the MC. Blue lines are the best-fit colour excess profiles resulted from 100 randomly picked sample by the Monte Carlo analysis.}
    \label{fig:162_a}
\end{figure*}

In this work, we find a MC feature at the distance bin of 0.5 to 1.0\,kpc. However, few overlaps have been found between the MC and HB9. On the contrary, it shows an anti-correlated pattern (Fig.~\ref{fig:162_a}), which is similar to the case of SNR G182.4+4.3. The distance of the cloud is estimated as $d_0$ = $556^{+74}_{-6}$\,pc and $\delta E(G-K_{\rm S})=0.43_{-0.12}^{+0.04} \rm\,mag$.

\subsection{G206.9+2.3, PKS 0646+06}
G206.9+2.3 is an SNR close to the Monoceros Nebula with a bright radio shell in the northwest. \citet{Graham1982A&A...109..145G} estimate the distance of the SNR to be from 3 to 5\,kpc. They obtain a spectral index of $\alpha=-0.45\pm0.03$, which is confirmed by \citet{Gao2011A&A...532A.144G} ($\alpha=-0.47\pm0.04$). According to the X-ray and optical studies, the SNR is probably evolving in low-density environment \citep{Leahy1986A&A...156..191L, Ambrocio2014RMxAA..50..323A}. Based on a large CO map in the direction of G206.9+2.3, \citet{Su2017ApJ...836..211S} detected a molecular gas cavity in the region where the remnant located.

\begin{figure*}
    \includegraphics[width=17.5cm]{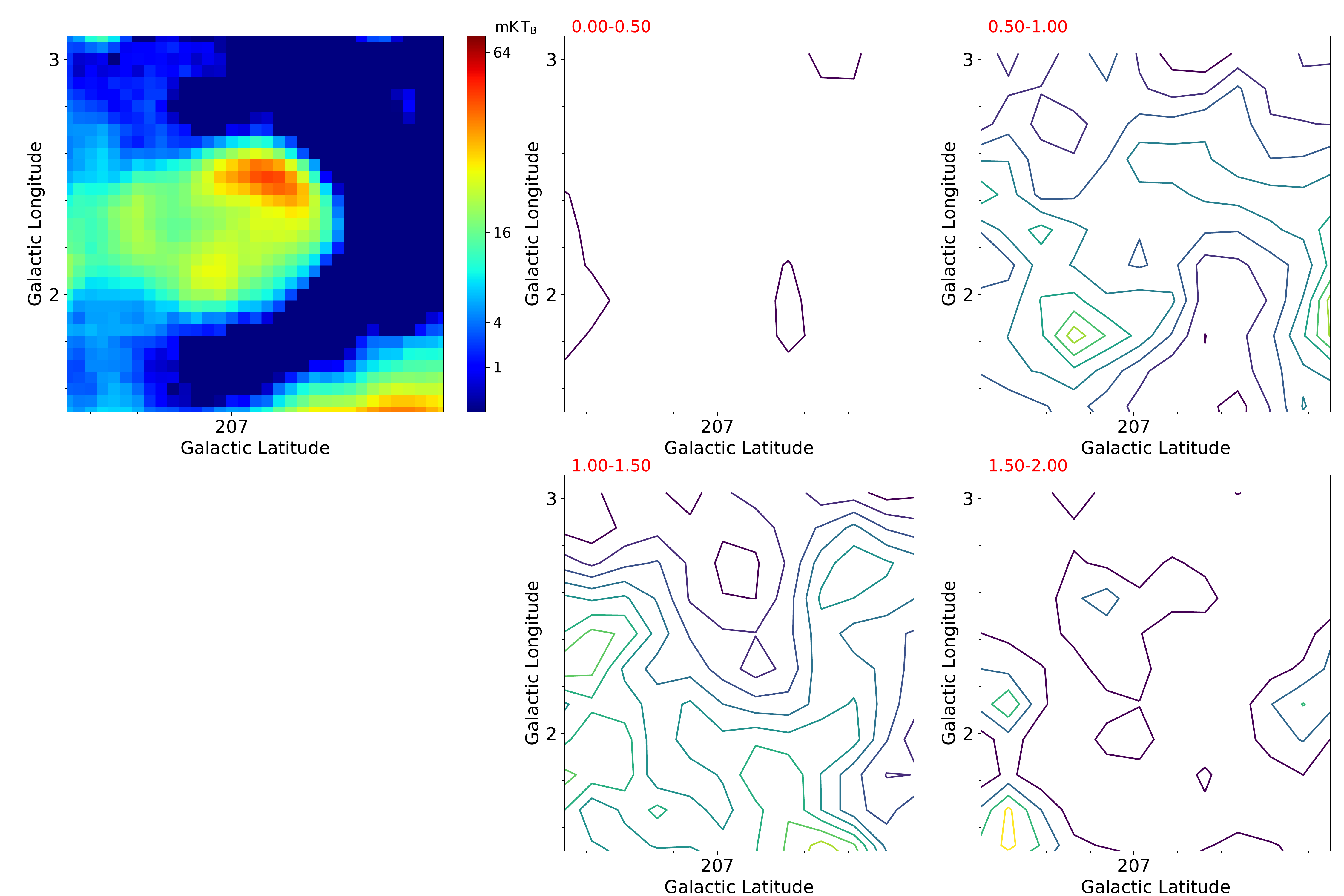}
    \caption{{\it Panel 1:} the morphologies of the 6\,cm radio emissions of G206.9+2.3. {\it Panel 2-5:} the colour excess distribution maps of G206.9+2.3 at distance bin 0.0-0.5\,kpc, 0.5-1.0\,kpc, 1.0-1.5\,kpc, 1.5-2.0\,kpc. The colour contours in the middle and right panels represent the distribution of the differential colour excesses, which vary from 0.08\,mag to 0.40\,mag with interval of 0.04\,mag. The colour map in the left panel represents the Urumqi 6\,cm radio image of G206.9+2.3.}
    \label{fig:173_a}
\end{figure*}

As shown in Fig.~\ref{fig:173_a}, the low colour excess values ($\leq 0.4 \rm\,mag$) in the region of G206.9+2.3 from our work agrees well with the low density and the lack of MCs suggested by \citet{Su2017ApJ...836..211S}. No obvious morphological correlation are found between the dust and CO observations.

\subsection{G156.2+5.7}
G156.2+5.7 is the first Galactic SNR discovered through its X-ray emission \citep{Pfeffermann1991A&A...246L..28P}. They conclude that the remnant is located in a region of very low interstellar density (0.01 atoms\,$\rm cm^{-3}$). It is one of the X-ray brightest and radio faintest SNRs known \citep{Reich1992A&A...256..214R, Xu2007A&A...470..969X}. Its radio morphology shows limb brightening along the northwest and southeast rim, which is typically seen in "barrel-shaped" SNRs \citep{Kesteven1987A&A...183..118K}. Based on the measurement of the expansion velocity of the SNR, \citet{Katsuda2016ApJ...826..108K} estimate the distance of the SNR as $d$ $\ge$1.7\,kpc. In the NE filament rim, spatial coincidence between shocked radiative emission and a dusty interstellar cloud is found \citep{Gerardy2007MNRAS.376..929G}. That may suggest that the remnant is colliding with the cloud. However, there is no clear evidence for that. There is a `hole' in the southwest side of the X-ray emission map of G156.2+5.7. \citet{Gerardy2007MNRAS.376..929G} believe the observed coincidences between X-ray emission holes and dusty interstellar clouds indicate there are clouds in the foreground of the remnant leaving X-ray absorption shadows on its X-ray image.

We have detected strong MC features, which have colour excess up to $E(G-K_{\rm S})$ = 0.97\,mag. The MC overlaps with the 6\,cm radio emission of the SNR on the west side. However, there is no obvious coincidence between their morphologies. The relatively close distance (about 0.7\,kpc) and high dust density we find are not consistent with the previous works. We suggest that it is a foreground MC.

\begin{figure*}
    \includegraphics[width=17.5cm]{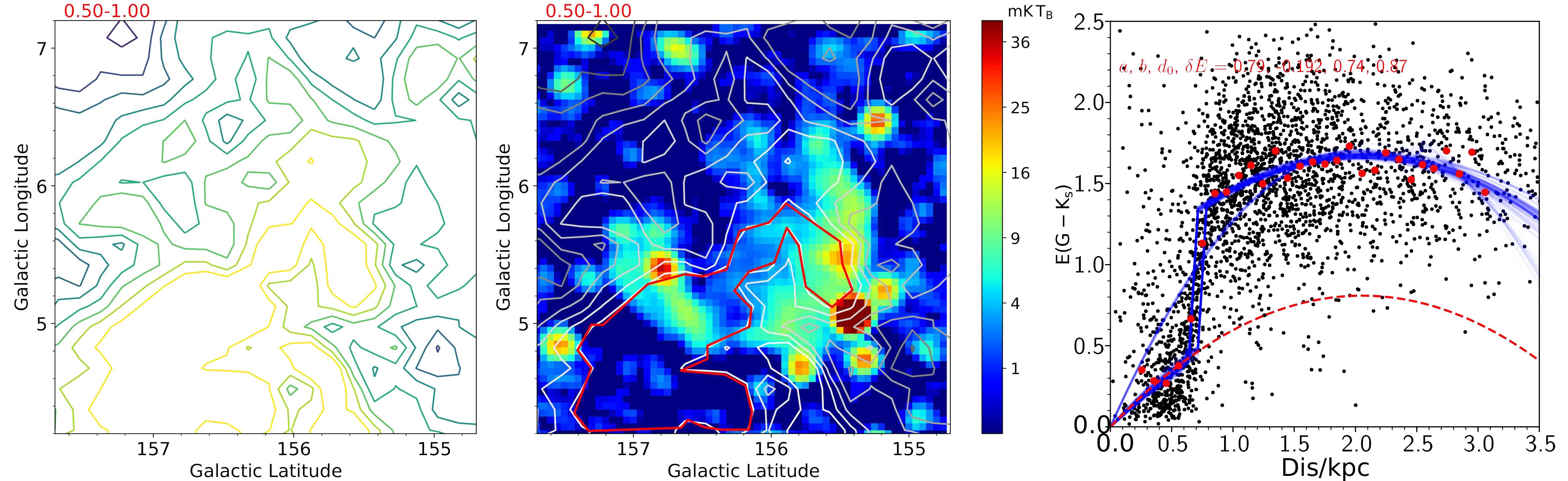}
    \caption{{\it Left:} the colour excess distribution map of G156.2+5.7 at distance bin 0.5-1.0\,kpc. {\it Middle: } the reddening region selected from the map, with the morphologies of the 6\,cm radio emissions. {\it Right:} the colour excess profile fitting for the MC. The colour contours in the left panel and the grey-scale contours in the middle panel represent the distribution of the differential colour excesses, which vary from 0.28\,mag to 0.97\,mag with interval of 0.14\,mag. The colour map in the middle panel represents the Urumqi 6\,cm radio image of G156.2+5.7. The red polygon in the middle panel represent the corresponding region of MC we selected. Red points are median values in each distance bin. Red dashed lines are extinction curve without the contribution of the MC. Blue lines are the best-fit colour excess profiles resulted from 100 randomly picked sample by the Monte Carlo analysis.}
    \label{fig:160_a}
\end{figure*}

\subsection{G166.0+4.3, VRO 42.05.01}
\begin{figure*}
    \includegraphics[width=17.5cm]{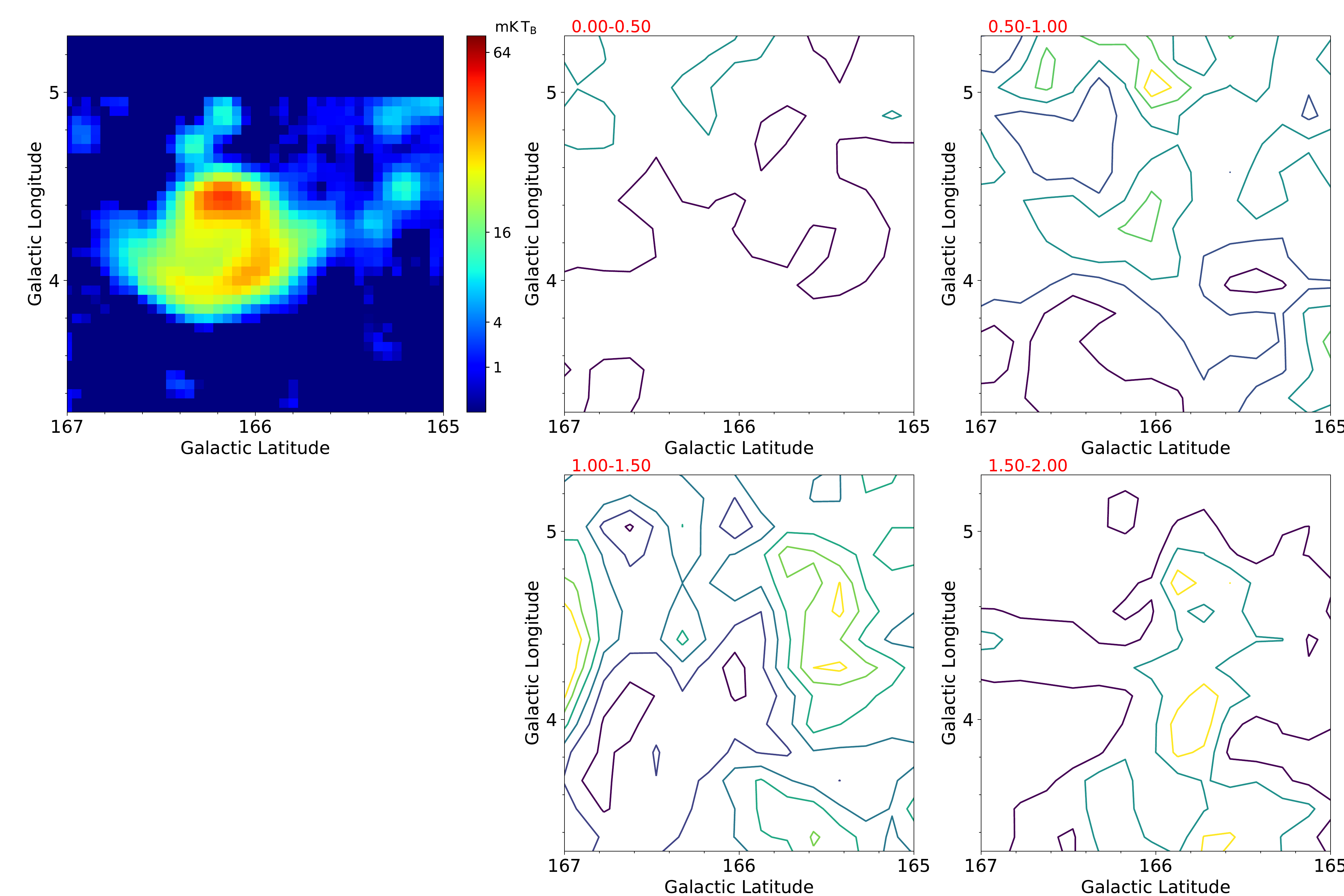}
    \caption{{\it Panel 1:} the morphologies of the 6\,cm radio emissions of G166.0+4.3. {\it Panel 2-5:} the colour excess distribution maps of G166.0+4.3 at distance bin 0.0-0.5\,kpc, 0.5-1.0\,kpc, 1.0-1.5\,kpc, 1.5-2.0\,kpc. The colour contours in the middle and right panels represent the distribution of the differential colour excesses, which vary from 0.11\,mag to 0.40\,mag with interval of 0.06\,mag. The colour map in the left panel represents the Urumqi 6\,cm radio image of G166.0+4.3.}
    \label{fig:163_a}
\end{figure*}

G166.0+4.3 has an unusual shape consisting of two radio shells with significantly different radius. \citet{Bocchino2009A&A...498..139B} classify G166.0+4.3 as an MM (Mixed-morphology) SNR. This unusual morphology and different radius may be explained by the different gas environments on both sides. The spectral index from 408 MHz to 1420 MHz is $-0.36\pm0.10$, calculated from flux densities without background point sources\citep{Leahy2005A&A...440..929L,Tian2006A&A...451..991T}. \citet{Kothes2006A&A...457.1081K} present an overall radio spectral index of $\alpha~=~-0.37\pm0.03$. Its distance is estimated to be $5.0\pm0.5$\,kpc \citep{Landecker1989MNRAS.237..277L}. The SNR is likely associated with molecular clouds \citep{Lazendic2006ApJ...647..350L,Ouchi2017arXiv170407455O}. However, we are not able to find any MCs that have spatial correlation with the SNR (Fig.~\ref{fig:163_a}). Considering the distance of the SNR mentioned above, we may not be able to trace the MC which is correlated with the SNR as our colour excess map is only complete at distance of 3-4\,kpc.

\subsection{G178.2-4.2}
\citet{Gao2011A&A...532A.144G} have discovered G178.2-4.2 as an SNR with strongly polarized emission detected along its northern shell. It shows a round morphology with a prominent shell on its north. The radio source 3C139.2 is located in the center of G178.2-4.2, but it has no relation to the remnant. After excluding the point source, the structure of its shells is rather straight-forward.

\begin{figure*}
    \includegraphics[width=17.5cm]{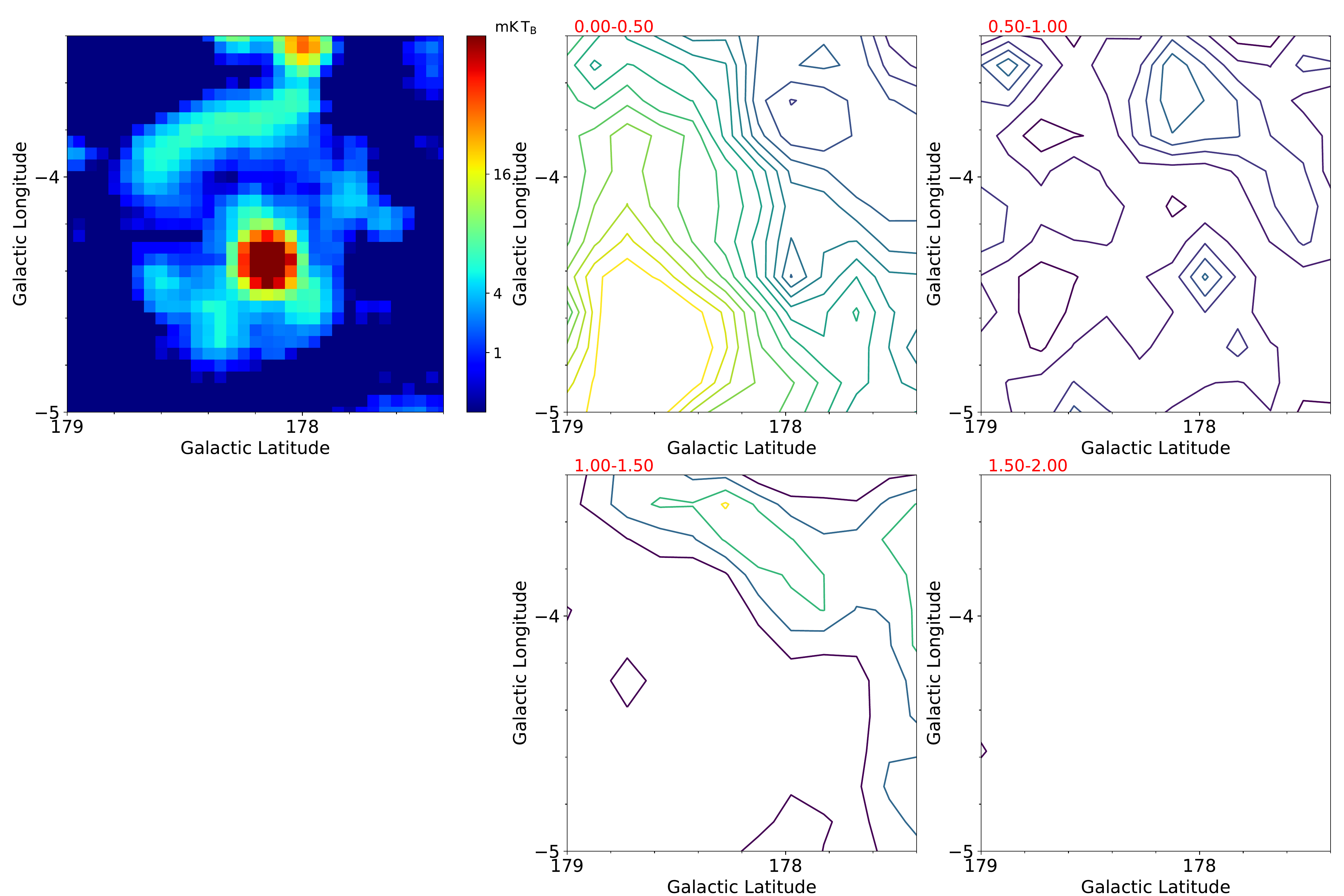}
    \caption{{\it Panel 1:} the morphologies of the 6\,cm radio emissions of G178.2-4.2. {\it Panel 2-5:} the colour excess distribution maps of G178.2-4.2 at distance bin 0.0-0.5\,kpc, 0.5-1.0\,kpc, 1.0-1.5\,kpc, 1.5-2.0\,kpc. The colour contours in the middle and right panels represent the distribution of the differential colour excesses, which vary from 0.12\,mag to 1.06\,mag with interval of 0.06\,mag. The colour map in the left panel represents the Urumqi 6\,cm radio image of G178.2-4.2.}
    \label{fig:164_a}
\end{figure*}

In the direction of G178.2-4.2, there is a strong MC feature with colour excess $E(G-K_{\rm S})$ up to 1.06\,mag at the closest  distance bin (0-0.5\,kpc). We are not able to find any MC features beyond 1.0\,kpc (Fig.~\ref{fig:164_a}). The morphology of the MC at the distance bin 0-0.5\,kpc has no spatial corrrelation with the SNR. 

\subsection{G179.0+2.6}
\begin{figure*}
    \includegraphics[width=17.5cm]{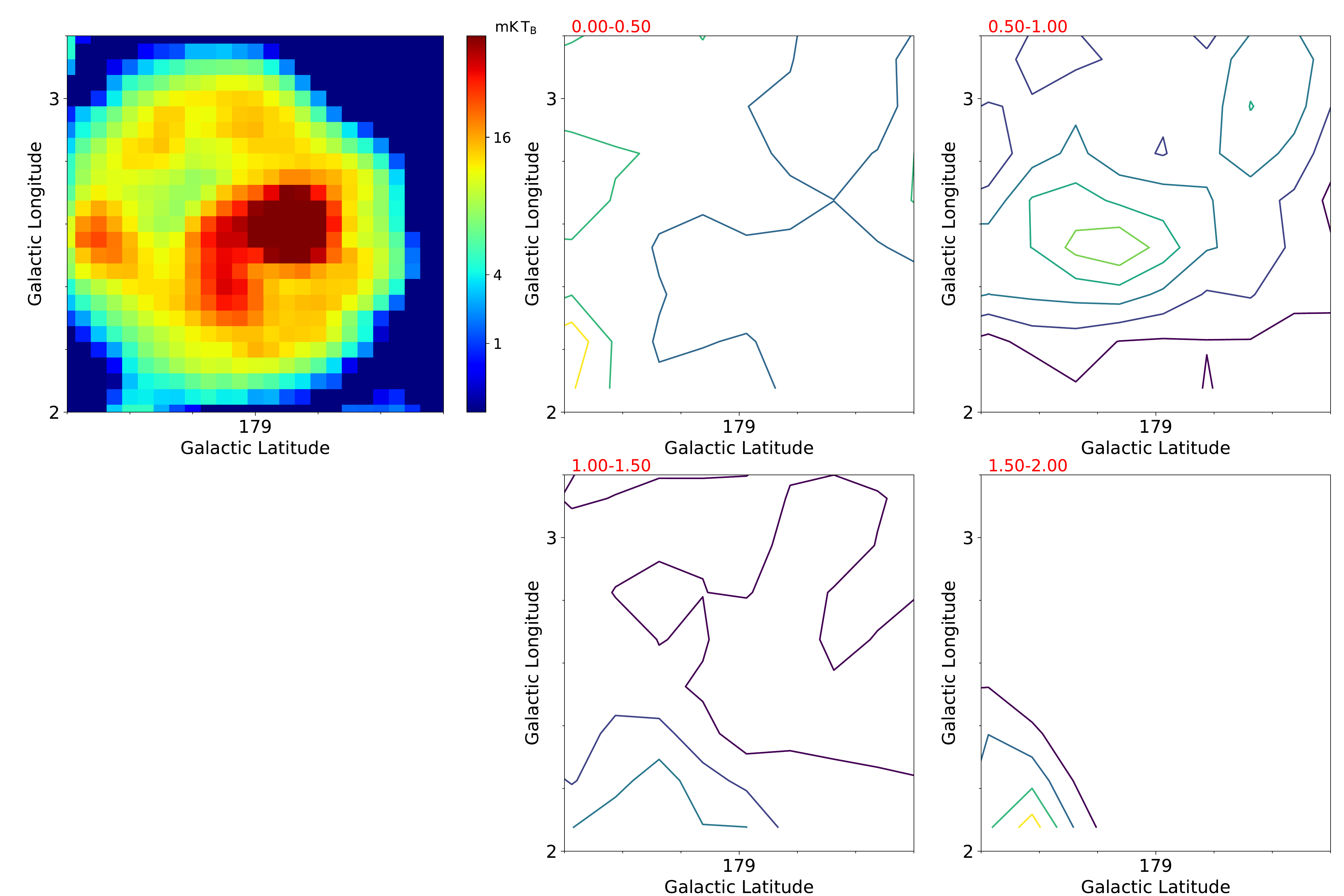}
    \caption{{\it Panel 1:} the morphologies of the 6\,cm radio emissions of G179.0+2.6. {\it Panel 2-5:} the colour excess distribution maps of G179.0+2.6 at distance bin 0.0-0.5\,kpc, 0.5-1.0\,kpc, 1.0-1.5\,kpc, 1.5-2.0\,kpc. The colour contours in the middle and right panels represent the distribution of the differential colour excesses, which vary from 0.11\,mag to 0.40\,mag with interval of 0.06\,mag. The colour map in the left panel represents the Urumqi 6\,cm radio image of G179.0+2.6.}
    \label{fig:165_a}
\end{figure*}

G179.0+2.6 is a faint shell-type SNR firstly identified and studied by \citet{Fuerst1986A&A...154..303F}. \citet{Fuerst1986A&A...154..303F} and \citet{Fuerst1989A&A...223...66F} discount the contribution of three extragalactic bright radio spots. In addition, \citet{Gao2011A&A...532A.144G} improve it by removing the flux contribution of 15 point-like sources and obtain an overall spectral index for the SNR G179.0+2.6 of $\alpha=-0.45\pm0.11$. We are not able to detect any MC that shows correlated features with the remnant (Fig.~\ref{fig:165_a}).

\section{Discussion and summary}
The 3D dust extinction mapping method can provide distance information to the SNRs interacting with MCs. This method is straight-forward, independent of and free from any influences of other assumptions. The morphology of the MCs identified can also serve as an indirect evidence of any potential interaction between SNRs and MCs. However, due to the limited resolution of our colour excess map and the completeness distance limit, we are not able to identify the MCs that associate with SNRs of small apparent diameter or SNRS at large distance from the Sun.

\begin{figure}
    \includegraphics[width=8cm]{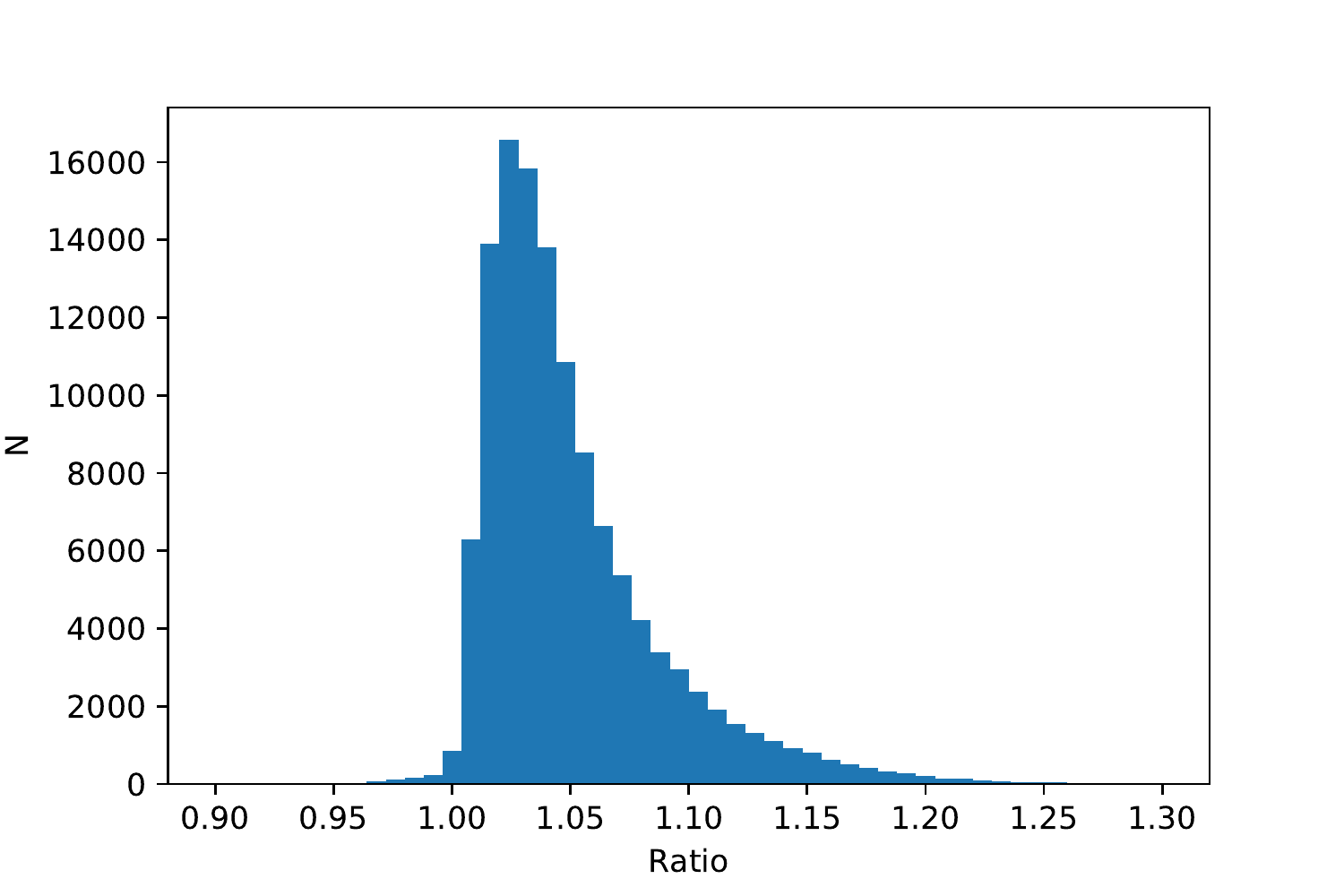}
    \caption{Ratio between distance directly transferred from the parallax of Gaia DR2 and the distance from the work of \citet{Bailer-Jones2018AJ....156...58B}}
    \label{fig:disratio}
\end{figure}

In this work, we have built 3D dust distribution maps toward 12 SNRs in the Galactic anticentre: G152.4-2.1, G156.2+5.7,  G160.9+2.6, G166.0+4.3, G178.2-4.2, G179.0+2.6, G182.4+4.3, G189.1+3.0, G190.9-2.2, G205.5+0.5, G206.9+2.3,  G213.0-0.6. Based on the maps, we have identified MCs which are potentially interacting with these SNRs by searching correlations of morphologies between the MCs and the 6\,cm radio emission of the SNR. Once any MCs are identified to have spatial correlation with the SNR, distances of the MCs (and the corresponding SNRs) are then determined by fitting the colour excess profile of the stars in the region of the MCs. We have found MCs which are spatially correlated with the radio extents for four SNRs: G189.1+3.0 (IC443),  G190.9-2.2, G205.5+0.5 (Monoceros Nebula), and G213.0-0.6. Accurate distances are presented for these SNRs. For three other SNRs (G182.4+4.3, G152.4-2.1, and G206.9+2.3), MCs that may potentially be related to the SNRs are found. These clouds are located near the remnants and have some features partly correlated or anti-correlated with the morphology of the radio emission of the corresponding SNRs. We provide estimations of distances for them. For the remaining remnants, no MCs with any obvious correlations with the SNRs are detected. Our results is summarized in Table.~\ref{tab: results}. The distances we determine are mostly consistent with previous works. However, the distance can be slightly underestimated due to the distance data we use. The distance of \citet{Bailer-Jones2018AJ....156...58B} use a method which is optimized for main sequence stars in long line of sight with normal extinction. Errors for lines of sight involving MCs may be more uncertain. We compare the distance transferred directly from Gaia DR2's parallax data with the distance we use for Monoceros nebula. As shown in Fig.~\ref{fig:disratio}, on average the distance directly from parallax is 5\% further. For individual stars, this can go up to 20\%.

In the future, this method can be further applied to all SNRs, especially those having certain interactions with MCs. As for those SNRs which have not found conclusive proof of interactions, we can use this method to test the possibility of interactions and obtain possible distances.

The whole reddening maps of 12 SNR are available online, containing reddening contours in all distance intervals of each SNR.

\begin{table*}
    \centering
    \caption{Distances of the SNRs resulted from this work.}
    \label{tab: results}
    \resizebox{\textwidth}{!}{
    \begin{tabular}{ccccccccc} 
        \hline
        No. & l & b & Size &  Distance & Distance & Name(s) & Physical Contact& Physical Contact\\
         & & & (arcmin)& This Work(kpc) & Previous(kpc) & with MCs in this work & with MCs in literature \\
        \hline
        1 & 189.1 & 3 & 45 & $1.73^{+0.13}_{-0.09}$  & 0.7-2 & IC443, 3C157 & Yes & OH maser, CO ratio,\\
         & & & &   & &   &  & H2, molecular MA \& LB\\
        2 & 190.9 & -2.2 & 70x60 & $1.03^{+0.02}_{-0.08}$ & $\rm 1.0\pm0.3$ & G190.9-2.2 & Yes & \\
        3 & 205.5 & 0.5 & 220 & $0.93^{+0.05}_{-0.08}/1.26^{+0.09}_{-0.10}$  & 0.8/1.6 & Monoceros Nebula & Yes & CO RC\\
        4 & 213 & -0.6 & 160x140? & $1.15^{+0.08}_{-0.08}$  & $\sim 1$ & G213-0.6 & Yes\\
        5 & 182.4 & 4.3 & 50 & $\sim$1.1  & $\geq 3$ & G182.4+4.3 & Possible\\
        6 & 152.4 & -2.1 & 100x95 & $\leq 1.0$ & $\sim 1.1$ & G152.4-2.1 & Possible \\
        7 & 160.9 & 2.6 & 140x120 & $\sim$0.6  & $0.8\pm0.4$ & HB9 & Possible & CO RC\\
        8 & 206.9 & 2.3 & 60x40 &   & 3-5\ & PKS 0646+06 & Not Found & \\
        9 & 156.2 & 5.7 & 110 &   & $\ge$1.7 & G156.2+5.7 &Not Found & \\
        10 & 166 & 4.3 & 55x35 &   & $5.0\pm0.5$ & VRO 42.05.01 &Not Found & CO RC\\
        11 & 178.2 & -4.2 & 72x62 &   & None & G178.2-4.2 &Not Found & \\
        12 & 179 & 2.6 & 70 &   & None & G179+2.6 &Not Found & \\

        \hline
    \end{tabular}}
\end{table*}

\section*{Acknowledgements}
This work is supported by the China Scholarship Council (No.201706040320) for 2 years' study at the University of Manchester.
BQC is supported by National Natural Science Foundation of China 11803029, U1531244, 11833006 and U1731308 and Yunnan University grant No.~C176220100007. BWJ is supported by NSFC through Project 11533002.

This work presents results from the European Space Agency (ESA) space mission Gaia. Gaia data are being processed by the Gaia Data Processing and Analysis Consortium (DPAC). Funding for the DPAC is provided by national institutions, in particular the institutions participating in the Gaia MultiLateral Agreement (MLA). The Gaia mission website is https://www.cosmos.esa.int/gaia. The Gaia archive website is https://archives.esac.esa.int/gaia.




\bibliographystyle{mnras}
\bibliography{texbibfile} 

\begin{thebibliography}{}
\makeatletter
\relax
\def\mn@urlcharsother{\let\do\@makeother \do\$\do\&\do\#\do\^\do\_\do\%\do\~}
\def\mn@doi{\begingroup\mn@urlcharsother \@ifnextchar [ {\mn@doi@}
  {\mn@doi@[]}}
\def\mn@doi@[#1]#2{\def\@tempa{#1}\ifx\@tempa\@empty \href
  {http://dx.doi.org/#2} {doi:#2}\else \href {http://dx.doi.org/#2} {#1}\fi
  \endgroup}
\def\mn@eprint#1#2{\mn@eprint@#1:#2::\@nil}
\def\mn@eprint@arXiv#1{\href {http://arxiv.org/abs/#1} {{\tt arXiv:#1}}}
\def\mn@eprint@dblp#1{\href {http://dblp.uni-trier.de/rec/bibtex/#1.xml}
  {dblp:#1}}
\def\mn@eprint@#1:#2:#3:#4\@nil{\def\@tempa {#1}\def\@tempb {#2}\def\@tempc
  {#3}\ifx \@tempc \@empty \let \@tempc \@tempb \let \@tempb \@tempa \fi \ifx
  \@tempb \@empty \def\@tempb {arXiv}\fi \@ifundefined
  {mn@eprint@\@tempb}{\@tempb:\@tempc}{\expandafter \expandafter \csname
  mn@eprint@\@tempb\endcsname \expandafter{\@tempc}}}

\bibitem[\protect\citeauthoryear{{Ambrocio-Cruz}, {Rosado}, {Le Coarer},
  {Bernal}  \& {Guti{\'e}rrez}}{{Ambrocio-Cruz}
  et~al.}{2014}]{Ambrocio2014RMxAA..50..323A}
{Ambrocio-Cruz} P.,  {Rosado} M.,  {Le Coarer} E.,  {Bernal} A.,
  {Guti{\'e}rrez} L.,  2014, \rmxaa, \href
  {http://adsabs.harvard.edu/abs/2014RMxAA..50..323A} {50, 323}

\bibitem[\protect\citeauthoryear{{Bailer-Jones}, {Rybizki}, {Fouesneau},
  {Mantelet}  \& {Andrae}}{{Bailer-Jones}
  et~al.}{2018}]{Bailer-Jones2018AJ....156...58B}
{Bailer-Jones} C.~A.~L.,  {Rybizki} J.,  {Fouesneau} M.,  {Mantelet} G.,
  {Andrae} R.,  2018, \mn@doi [\aj] {10.3847/1538-3881/aacb21}, \href
  {https://ui.adsabs.harvard.edu/#abs/2018AJ....156...58B} {156, 58}

\bibitem[\protect\citeauthoryear{{Bocchino}, {Miceli}  \& {Troja}}{{Bocchino}
  et~al.}{2009}]{Bocchino2009A&A...498..139B}
{Bocchino} F.,  {Miceli} M.,   {Troja} E.,  2009, \mn@doi [\aap]
  {10.1051/0004-6361/200810742}, \href
  {http://adsabs.harvard.edu/abs/2009A%26A...498..139B} {498, 139}

\bibitem[\protect\citeauthoryear{{Braun} \& {Strom}}{{Braun} \&
  {Strom}}{1986}]{Braun1986A&A...164..193B}
{Braun} R.,  {Strom} R.~G.,  1986, \aap, \href
  {http://ads.bao.ac.cn/abs/1986A%26A...164..193B} {164, 193}

\bibitem[\protect\citeauthoryear{{Case} \& {Bhattacharya}}{{Case} \&
  {Bhattacharya}}{1998}]{Case1998ApJ...504..761C}
{Case} G.~L.,  {Bhattacharya} D.,  1998, \mn@doi [\apj] {10.1086/306089}, \href
  {http://adsabs.harvard.edu/abs/1998ApJ...504..761C} {504, 761}

\bibitem[\protect\citeauthoryear{{Chen}, {Jiang}, {Zhou}, {Su}, {Zhou}, {Li}
  \& {Zhang}}{{Chen} et~al.}{2014a}]{Chen2014IAUS}
{Chen} Y.,  {Jiang} B.,  {Zhou} P.,  {Su} Y.,  {Zhou} X.,  {Li} H.,   {Zhang}
  X.,  2014a, in {Ray} A.,  {McCray} R.~A.,  eds,  IAU Symposium Vol. 296,
  Supernova Environmental Impacts. pp 170--177 (\mn@eprint {arXiv}
  {1304.5367}), \mn@doi{10.1017/S1743921313009423}

\bibitem[\protect\citeauthoryear{{Chen} et~al.,}{{Chen}
  et~al.}{2014b}]{Chen2014MNRAS}
{Chen} B.-Q.,  et~al., 2014b, \mn@doi [\mnras] {10.1093/mnras/stu1192}, \href
  {http://adsabs.harvard.edu/abs/2014MNRAS.443.1192C} {443, 1192}

\bibitem[\protect\citeauthoryear{{Chen}, {Liu}, {Yuan}, {Huang}  \&
  {Xiang}}{{Chen} et~al.}{2015}]{Chen2015MNRAS.448.2187C}
{Chen} B.-Q.,  {Liu} X.-W.,  {Yuan} H.-B.,  {Huang} Y.,   {Xiang} M.-S.,  2015,
  \mn@doi [\mnras] {10.1093/mnras/stv103}, \href
  {http://adsabs.harvard.edu/abs/2015MNRAS.448.2187C} {448, 2187}

\bibitem[\protect\citeauthoryear{{Chen} et~al.,}{{Chen}
  et~al.}{2017}]{Chen2017MNRAS.472.3924C}
{Chen} B.-Q.,  et~al., 2017, \mn@doi [\mnras] {10.1093/mnras/stx2287}, \href
  {http://adsabs.harvard.edu/abs/2017MNRAS.472.3924C} {472, 3924}

\bibitem[\protect\citeauthoryear{{Chen} et~al.,}{{Chen}
  et~al.}{2019}]{chen2019MNRAS.483.4277C}
{Chen} B.-Q.,  et~al., 2019, \mn@doi [\mnras] {10.1093/mnras/sty3341}, \href
  {http://adsabs.harvard.edu/abs/2019MNRAS.483.4277C} {483, 4277}

\bibitem[\protect\citeauthoryear{{Claussen}, {Frail}, {Goss}  \&
  {Gaume}}{{Claussen} et~al.}{1997}]{Claussen1997ApJ...489..143C}
{Claussen} M.~J.,  {Frail} D.~A.,  {Goss} W.~M.,   {Gaume} R.~A.,  1997,
  \mn@doi [\apj] {10.1086/304784}, \href
  {http://ads.bao.ac.cn/abs/1997ApJ...489..143C} {489, 143}

\bibitem[\protect\citeauthoryear{{Cornett}, {Chin}  \& {Knapp}}{{Cornett}
  et~al.}{1977}]{Cornett1977A&A....54..889C}
{Cornett} R.~H.,  {Chin} G.,   {Knapp} G.~R.,  1977, \aap, \href
  {http://ads.bao.ac.cn/abs/1977A%26A....54..889C} {54, 889}

\bibitem[\protect\citeauthoryear{{Davies}, {Elliott}, {Goudis}, {Meaburn}  \&
  {Tebbutt}}{{Davies} et~al.}{1978}]{Davies1978A&AS...31..271D}
{Davies} R.~D.,  {Elliott} K.~H.,  {Goudis} C.,  {Meaburn} J.,   {Tebbutt}
  N.~J.,  1978, \aaps, \href
  {http://adsabs.harvard.edu/abs/1978A%26AS...31..271D} {31, 271}

\bibitem[\protect\citeauthoryear{{Denoyer}}{{Denoyer}}{1978}]{Denoyer1978MNRAS.183..187D}
{Denoyer} L.~K.,  1978, \mn@doi [\mnras] {10.1093/mnras/183.2.187}, \href
  {http://ads.bao.ac.cn/abs/1978MNRAS.183..187D} {183, 187}

\bibitem[\protect\citeauthoryear{{Denoyer}}{{Denoyer}}{1979a}]{Denoyer1979ApJ...228L..41D}
{Denoyer} L.~K.,  1979a, \mn@doi [\apjl] {10.1086/182899}, \href
  {http://ads.bao.ac.cn/abs/1979ApJ...228L..41D} {228, L41}

\bibitem[\protect\citeauthoryear{{Denoyer}}{{Denoyer}}{1979b}]{Denoyer1979ApJ...232L.165D}
{Denoyer} L.~K.,  1979b, \mn@doi [\apjl] {10.1086/183057}, \href
  {http://ads.bao.ac.cn/abs/1979ApJ...232L.165D} {232, L165}

\bibitem[\protect\citeauthoryear{{Dickman}, {Snell}, {Ziurys}  \&
  {Huang}}{{Dickman} et~al.}{1992}]{Dickman1992ApJ...400..203D}
{Dickman} R.~L.,  {Snell} R.~L.,  {Ziurys} L.~M.,   {Huang} Y.-L.,  1992,
  \mn@doi [\apj] {10.1086/171987}, \href
  {http://ads.bao.ac.cn/abs/1992ApJ...400..203D} {400, 203}

\bibitem[\protect\citeauthoryear{{Fiasson}, {Hinton}, {Gallant}, {Marcowith},
  {Reimer}  \& {Rowell}}{{Fiasson} et~al.}{2008}]{Fiasson2008ICRC....2..719F}
{Fiasson} A.,  {Hinton} J.~A.,  {Gallant} Y.,  {Marcowith} A.,  {Reimer} O.,
  {Rowell} G.,  2008, International Cosmic Ray Conference, \href
  {http://adsabs.harvard.edu/abs/2008ICRC....2..719F} {2, 719}

\bibitem[\protect\citeauthoryear{{Foster}, {Cooper}, {Reich}, {Kothes}  \&
  {West}}{{Foster} et~al.}{2013}]{Foster2013A&A...549A.107F}
{Foster} T.~J.,  {Cooper} B.,  {Reich} W.,  {Kothes} R.,   {West} J.,  2013,
  \mn@doi [\aap] {10.1051/0004-6361/201220362}, \href
  {http://adsabs.harvard.edu/abs/2013A%26A...549A.107F} {549, A107}

\bibitem[\protect\citeauthoryear{{Fountain}, {Gary}  \& {Odell}}{{Fountain}
  et~al.}{1979}]{Fountain1979ApJ...229..971F}
{Fountain} W.~F.,  {Gary} G.~A.,   {Odell} C.~R.,  1979, \mn@doi [\apj]
  {10.1086/157031}, \href {http://adsabs.harvard.edu/abs/1979ApJ...229..971F}
  {229, 971}

\bibitem[\protect\citeauthoryear{{Fuerst} \& {Reich}}{{Fuerst} \&
  {Reich}}{1986}]{Fuerst1986A&A...154..303F}
{Fuerst} E.,  {Reich} W.,  1986, \aap, \href
  {http://adsabs.harvard.edu/abs/1986A%26A...154..303F} {154, 303}

\bibitem[\protect\citeauthoryear{{Fuerst}, {Reich}, {Kuehr}  \&
  {Stickel}}{{Fuerst} et~al.}{1989}]{Fuerst1989A&A...223...66F}
{Fuerst} E.,  {Reich} W.,  {Kuehr} H.,   {Stickel} M.,  1989, \aap, \href
  {http://adsabs.harvard.edu/abs/1989A%26A...223...66F} {223, 66}

\bibitem[\protect\citeauthoryear{{Gao} et~al.,}{{Gao}
  et~al.}{2010}]{Gao2010A&A...515A..64G}
{Gao} X.~Y.,  et~al., 2010, \mn@doi [\aap] {10.1051/0004-6361/200913793}, \href
  {http://adsabs.harvard.edu/abs/2010A%26A...515A..64G} {515, A64}

\bibitem[\protect\citeauthoryear{{Gao}, {Sun}, {Han}, {Reich}, {Reich}  \&
  {Wielebinski}}{{Gao} et~al.}{2011}]{Gao2011A&A...532A.144G}
{Gao} X.~Y.,  {Sun} X.~H.,  {Han} J.~L.,  {Reich} W.,  {Reich} P.,
  {Wielebinski} R.,  2011, \mn@doi [\aap] {10.1051/0004-6361/201117179}, \href
  {http://adsabs.harvard.edu/abs/2011A%26A...532A.144G} {532, A144}

\bibitem[\protect\citeauthoryear{{Gerardy} \& {Fesen}}{{Gerardy} \&
  {Fesen}}{2007}]{Gerardy2007MNRAS.376..929G}
{Gerardy} C.~L.,  {Fesen} R.~A.,  2007, \mn@doi [\mnras]
  {10.1111/j.1365-2966.2007.11494.x}, \href
  {http://adsabs.harvard.edu/abs/2007MNRAS.376..929G} {376, 929}

\bibitem[\protect\citeauthoryear{{Goodman}, {Pineda}  \& {Schnee}}{{Goodman}
  et~al.}{2009}]{Goodman2009ApJ...692...91G}
{Goodman} A.~A.,  {Pineda} J.~E.,   {Schnee} S.~L.,  2009, \mn@doi [\apj]
  {10.1088/0004-637X/692/1/91}, \href
  {http://adsabs.harvard.edu/abs/2009ApJ...692...91G} {692, 91}

\bibitem[\protect\citeauthoryear{{Graham}, {Haslam}, {Salter}  \&
  {Wilson}}{{Graham} et~al.}{1982}]{Graham1982A&A...109..145G}
{Graham} D.~A.,  {Haslam} C.~G.~T.,  {Salter} C.~J.,   {Wilson} W.~E.,  1982,
  \aap, \href {http://adsabs.harvard.edu/abs/1982A%26A...109..145G} {109, 145}

\bibitem[\protect\citeauthoryear{{Green}}{{Green}}{2017}]{GreenCatalog}
{Green} D.,  2017, A Catalogue of Galactic Supernova Remnants (2017 June
  version), \url {http://www.mrao.cam.ac.uk/surveys/snrs/}

\bibitem[\protect\citeauthoryear{{Green} et~al.,}{{Green}
  et~al.}{2015}]{Green2015ApJ...810...25G}
{Green} G.~M.,  et~al., 2015, \mn@doi [\apj] {10.1088/0004-637X/810/1/25},
  \href {http://adsabs.harvard.edu/abs/2015ApJ...810...25G} {810, 25}

\bibitem[\protect\citeauthoryear{{Guseinov}, {Ankay}, {Sezer}  \&
  {Tagieva}}{{Guseinov} et~al.}{2003}]{Guseinov2003A&AT}
{Guseinov} O.~H.,  {Ankay} A.,  {Sezer} A.,   {Tagieva} S.~O.,  2003, \mn@doi
  [Astronomical and Astrophysical Transactions] {10.1080/1055679021000034160},
  \href {http://adsabs.harvard.edu/abs/2003A%26AT...22..273G} {22, 273}

\bibitem[\protect\citeauthoryear{{Hewitt}, {Yusef-Zadeh}, {Wardle}, {Roberts}
  \& {Kassim}}{{Hewitt} et~al.}{2006}]{Hewitt2006ApJ...652.1288H}
{Hewitt} J.~W.,  {Yusef-Zadeh} F.,  {Wardle} M.,  {Roberts} D.~A.,   {Kassim}
  N.~E.,  2006, \mn@doi [\apj] {10.1086/508331}, \href
  {http://ads.bao.ac.cn/abs/2006ApJ...652.1288H} {652, 1288}

\bibitem[\protect\citeauthoryear{{Huang} \& {Thaddeus}}{{Huang} \&
  {Thaddeus}}{1986}]{Huang1986ApJ}
{Huang} Y.-L.,  {Thaddeus} P.,  1986, \mn@doi [\apj] {10.1086/164649}, \href
  {http://adsabs.harvard.edu/abs/1986ApJ...309..804H} {309, 804}

\bibitem[\protect\citeauthoryear{{Huang}, {Dickman}  \& {Snell}}{{Huang}
  et~al.}{1986}]{Huang1986ApJ...302L..63H}
{Huang} Y.-L.,  {Dickman} R.~L.,   {Snell} R.~L.,  1986, \mn@doi [\apjl]
  {10.1086/184638}, \href {http://ads.bao.ac.cn/abs/1986ApJ...302L..63H} {302,
  L63}

\bibitem[\protect\citeauthoryear{{Jaffe}, {Bhattacharya}, {Dixon}  \&
  {Zych}}{{Jaffe} et~al.}{1997}]{Jaffe1997ApJ...484L.129J}
{Jaffe} T.~R.,  {Bhattacharya} D.,  {Dixon} D.~D.,   {Zych} A.~D.,  1997,
  \mn@doi [\apjl] {10.1086/310797}, \href
  {http://adsabs.harvard.edu/abs/1997ApJ...484L.129J} {484, L129}

\bibitem[\protect\citeauthoryear{{Jeong}, {Byun}, {Koo}, {Lee}, {Lee}  \&
  {Kang}}{{Jeong} et~al.}{2012}]{Jeong2012Ap&SS}
{Jeong} I.-G.,  {Byun} D.-Y.,  {Koo} B.-C.,  {Lee} J.-J.,  {Lee} J.-W.,
  {Kang} H.,  2012, \mn@doi [\apss] {10.1007/s10509-012-1196-1}, \href
  {http://adsabs.harvard.edu/abs/2012Ap%26SS.342..389J} {342, 389}

\bibitem[\protect\citeauthoryear{{Jiang}, {Chen}, {Wang}, {Su}, {Zhou},
  {Safi-Harb}  \& {DeLaney}}{{Jiang} et~al.}{2010}]{Jiang2010ApJ}
{Jiang} B.,  {Chen} Y.,  {Wang} J.,  {Su} Y.,  {Zhou} X.,  {Safi-Harb} S.,
  {DeLaney} T.,  2010, \mn@doi [\apj] {10.1088/0004-637X/712/2/1147}, \href
  {http://adsabs.harvard.edu/abs/2010ApJ...712.1147J} {712, 1147}

\bibitem[\protect\citeauthoryear{{Katsuda}, {Tanaka}, {Morokuma}, {Fesen}  \&
  {Milisavljevic}}{{Katsuda} et~al.}{2016}]{Katsuda2016ApJ...826..108K}
{Katsuda} S.,  {Tanaka} M.,  {Morokuma} T.,  {Fesen} R.,   {Milisavljevic} D.,
  2016, \mn@doi [\apj] {10.3847/0004-637X/826/2/108}, \href
  {http://adsabs.harvard.edu/abs/2016ApJ...826..108K} {826, 108}

\bibitem[\protect\citeauthoryear{{Kesteven} \& {Caswell}}{{Kesteven} \&
  {Caswell}}{1987}]{Kesteven1987A&A...183..118K}
{Kesteven} M.~J.,  {Caswell} J.~L.,  1987, \aap, \href
  {http://adsabs.harvard.edu/abs/1987A%26A...183..118K} {183, 118}

\bibitem[\protect\citeauthoryear{{Kothes}, {Furst}  \& {Reich}}{{Kothes}
  et~al.}{1998}]{Kothes1998A&A...331..661K}
{Kothes} R.,  {Furst} E.,   {Reich} W.,  1998, \aap, \href
  {http://adsabs.harvard.edu/abs/1998A%26A...331..661K} {331, 661}

\bibitem[\protect\citeauthoryear{{Kothes}, {Fedotov}, {Foster}  \&
  {Uyan{\i}ker}}{{Kothes} et~al.}{2006}]{Kothes2006A&A...457.1081K}
{Kothes} R.,  {Fedotov} K.,  {Foster} T.~J.,   {Uyan{\i}ker} B.,  2006, \mn@doi
  [\aap] {10.1051/0004-6361:20065062}, \href
  {http://adsabs.harvard.edu/abs/2006A%26A...457.1081K} {457, 1081}

\bibitem[\protect\citeauthoryear{{Landecker}, {Pineault}, {Routledge}  \&
  {Vaneldik}}{{Landecker} et~al.}{1989}]{Landecker1989MNRAS.237..277L}
{Landecker} T.~L.,  {Pineault} S.,  {Routledge} D.,   {Vaneldik} J.~F.,  1989,
  \mn@doi [\mnras] {10.1093/mnras/237.1.277}, \href
  {http://adsabs.harvard.edu/abs/1989MNRAS.237..277L} {237, 277}

\bibitem[\protect\citeauthoryear{{Lazendic} \& {Slane}}{{Lazendic} \&
  {Slane}}{2006}]{Lazendic2006ApJ...647..350L}
{Lazendic} J.~S.,  {Slane} P.~O.,  2006, \mn@doi [\apj] {10.1086/505380}, \href
  {http://adsabs.harvard.edu/abs/2006ApJ...647..350L} {647, 350}

\bibitem[\protect\citeauthoryear{{Leahy}}{{Leahy}}{1986}]{Leahy1986A&A...156..191L}
{Leahy} D.~A.,  1986, \aap, \href
  {http://adsabs.harvard.edu/abs/1986A%26A...156..191L} {156, 191}

\bibitem[\protect\citeauthoryear{{Leahy}}{{Leahy}}{2004}]{Leahy2004AJ....127.2277L}
{Leahy} D.~A.,  2004, \mn@doi [\aj] {10.1086/382241}, \href
  {http://adsabs.harvard.edu/abs/2004AJ....127.2277L} {127, 2277}

\bibitem[\protect\citeauthoryear{{Leahy} \& {Roger}}{{Leahy} \&
  {Roger}}{1991}]{Leahy1991AJ....101.1033L}
{Leahy} D.~A.,  {Roger} R.~S.,  1991, \mn@doi [\aj] {10.1086/115745}, \href
  {http://adsabs.harvard.edu/abs/1991AJ....101.1033L} {101, 1033}

\bibitem[\protect\citeauthoryear{{Leahy} \& {Tian}}{{Leahy} \&
  {Tian}}{2005}]{Leahy2005A&A...440..929L}
{Leahy} D.,  {Tian} W.,  2005, \mn@doi [\aap] {10.1051/0004-6361:20053392},
  \href {http://adsabs.harvard.edu/abs/2005A%26A...440..929L} {440, 929}

\bibitem[\protect\citeauthoryear{{Leahy} \& {Tian}}{{Leahy} \&
  {Tian}}{2007}]{Leahy2007A&A...461.1013L}
{Leahy} D.~A.,  {Tian} W.~W.,  2007, \mn@doi [\aap]
  {10.1051/0004-6361:20065895}, \href
  {http://adsabs.harvard.edu/abs/2007A%26A...461.1013L} {461, 1013}

\bibitem[\protect\citeauthoryear{{Leahy} \& {Tian}}{{Leahy} \&
  {Tian}}{2008}]{Leahy2008AJ}
{Leahy} D.~A.,  {Tian} W.~W.,  2008, \mn@doi [\aj]
  {10.1088/0004-6256/135/1/167}, \href
  {http://adsabs.harvard.edu/abs/2008AJ....135..167L} {135, 167}

\bibitem[\protect\citeauthoryear{{Lee}, {Koo}, {Yun}, {Stanimirovi{\'c}},
  {Heiles}  \& {Heyer}}{{Lee} et~al.}{2008}]{Lee2008AJ....135..796L}
{Lee} J.-J.,  {Koo} B.-C.,  {Yun} M.~S.,  {Stanimirovi{\'c}} S.,  {Heiles} C.,
   {Heyer} M.,  2008, \mn@doi [\aj] {10.1088/0004-6256/135/3/796}, \href
  {http://ads.bao.ac.cn/abs/2008AJ....135..796L} {135, 796}

\bibitem[\protect\citeauthoryear{{Odegard}}{{Odegard}}{1986}]{Odegard1986ApJ...301..813O}
{Odegard} N.,  1986, \mn@doi [\apj] {10.1086/163945}, \href
  {http://adsabs.harvard.edu/abs/1986ApJ...301..813O} {301, 813}

\bibitem[\protect\citeauthoryear{{Olbert}, {Clearfield}, {Williams}, {Keohane}
  \& {Frail}}{{Olbert} et~al.}{2001}]{Olbert2001ApJ...554L.205O}
{Olbert} C.~M.,  {Clearfield} C.~R.,  {Williams} N.~E.,  {Keohane} J.~W.,
  {Frail} D.~A.,  2001, \mn@doi [\apjl] {10.1086/321708}, \href
  {http://ads.bao.ac.cn/abs/2001ApJ...554L.205O} {554, L205}

\bibitem[\protect\citeauthoryear{{Ouchi} et~al.,}{{Ouchi}
  et~al.}{2017}]{Ouchi2017arXiv170407455O}
{Ouchi} M.,  et~al., 2017, preprint, \href
  {http://adsabs.harvard.edu/abs/2017arXiv170407455O} {} (\mn@eprint {arXiv}
  {1704.07455})

\bibitem[\protect\citeauthoryear{{Pavlovic}, {Dobardzic}, {Vukotic}  \&
  {Urosevic}}{{Pavlovic} et~al.}{2014}]{Pavlovic2014SerAJ.189...25P}
{Pavlovic} M.~Z.,  {Dobardzic} A.,  {Vukotic} B.,   {Urosevic} D.,  2014,
  \mn@doi [Serbian Astronomical Journal] {10.2298/SAJ1489025P}, \href
  {http://adsabs.harvard.edu/abs/2014SerAJ.189...25P} {189, 25}

\bibitem[\protect\citeauthoryear{{Pfeffermann}, {Aschenbach}  \&
  {Predehl}}{{Pfeffermann} et~al.}{1991}]{Pfeffermann1991A&A...246L..28P}
{Pfeffermann} E.,  {Aschenbach} B.,   {Predehl} P.,  1991, \aap, \href
  {http://adsabs.harvard.edu/abs/1991A%26A...246L..28P} {246, L28}

\bibitem[\protect\citeauthoryear{{Planck Collaboration} et~al.,}{{Planck
  Collaboration} et~al.}{2011}]{Planck2011A&A...536A..19P}
{Planck Collaboration} et~al., 2011, \mn@doi [\aap]
  {10.1051/0004-6361/201116479}, \href
  {http://adsabs.harvard.edu/abs/2011A%26A...536A..19P} {536, A19}

\bibitem[\protect\citeauthoryear{{Reich}, {Fuerst}  \& {Arnal}}{{Reich}
  et~al.}{1992}]{Reich1992A&A...256..214R}
{Reich} W.,  {Fuerst} E.,   {Arnal} E.~M.,  1992, \aap, \href
  {http://adsabs.harvard.edu/abs/1992A%26A...256..214R} {256, 214}

\bibitem[\protect\citeauthoryear{{Reich}, {Zhang}  \& {F{\"u}rst}}{{Reich}
  et~al.}{2003}]{Reich2003A&A...408..961R}
{Reich} W.,  {Zhang} X.,   {F{\"u}rst} E.,  2003, \mn@doi [\aap]
  {10.1051/0004-6361:20030939}, \href
  {http://adsabs.harvard.edu/abs/2003A%26A...408..961R} {408, 961}

\bibitem[\protect\citeauthoryear{{Shan}, {Zhu}, {Tian}, {Zhang}, {Zhang}, {Wu}
  \& {Yang}}{{Shan} et~al.}{2018}]{Shan2018ApJS}
{Shan} S.~S.,  {Zhu} H.,  {Tian} W.~W.,  {Zhang} M.~F.,  {Zhang} H.~Y.,  {Wu}
  D.,   {Yang} A.~Y.,  2018, \mn@doi [The Astrophysical Journal Supplement
  Series] {10.3847/1538-4365/aae07a}, \href
  {https://ui.adsabs.harvard.edu/\#abs/2018ApJS..238...35S} {238, 35}

\bibitem[\protect\citeauthoryear{{Snell}, {Hollenbach}, {Howe}, {Neufeld},
  {Kaufman}, {Melnick}, {Bergin}  \& {Wang}}{{Snell}
  et~al.}{2005}]{Snell2005ApJ...620..758S}
{Snell} R.~L.,  {Hollenbach} D.,  {Howe} J.~E.,  {Neufeld} D.~A.,  {Kaufman}
  M.~J.,  {Melnick} G.~J.,  {Bergin} E.~A.,   {Wang} Z.,  2005, \mn@doi [\apj]
  {10.1086/427231}, \href {http://ads.bao.ac.cn/abs/2005ApJ...620..758S} {620,
  758}

\bibitem[\protect\citeauthoryear{{Su} et~al.,}{{Su}
  et~al.}{2017}]{Su2017ApJ...836..211S}
{Su} Y.,  et~al., 2017, \mn@doi [\apj] {10.3847/1538-4357/aa5cb7}, \href
  {http://adsabs.harvard.edu/abs/2017ApJ...836..211S} {836, 211}

\bibitem[\protect\citeauthoryear{{Tian} \& {Leahy}}{{Tian} \&
  {Leahy}}{2006}]{Tian2006A&A...451..991T}
{Tian} W.~W.,  {Leahy} D.~A.,  2006, \mn@doi [\aap]
  {10.1051/0004-6361:20042431e}, \href
  {http://adsabs.harvard.edu/abs/2006A%26A...451..991T} {451, 991}

\bibitem[\protect\citeauthoryear{{Tian} \& {Leahy}}{{Tian} \&
  {Leahy}}{2012}]{Tian2012MNRAS}
{Tian} W.~W.,  {Leahy} D.~A.,  2012, \mn@doi [\mnras]
  {10.1111/j.1365-2966.2012.20491.x}, \href
  {http://adsabs.harvard.edu/abs/2012MNRAS.421.2593T} {421, 2593}

\bibitem[\protect\citeauthoryear{{Tian}, {Leahy}  \& {Wang}}{{Tian}
  et~al.}{2007}]{Tian2007A&A}
{Tian} W.~W.,  {Leahy} D.~A.,   {Wang} Q.~D.,  2007, \mn@doi [\aap]
  {10.1051/0004-6361:20077527}, \href
  {http://adsabs.harvard.edu/abs/2007A%26A...474..541T} {474, 541}

\bibitem[\protect\citeauthoryear{{Tian}, {Leahy}, {Haverkorn}  \&
  {Jiang}}{{Tian} et~al.}{2008}]{Tian2008ApJ}
{Tian} W.~W.,  {Leahy} D.~A.,  {Haverkorn} M.,   {Jiang} B.,  2008, \mn@doi
  [\apjl] {10.1086/589506}, \href
  {http://adsabs.harvard.edu/abs/2008ApJ...679L..85T} {679, L85}

\bibitem[\protect\citeauthoryear{{Torres}, {Romero}, {Dame}, {Combi}  \&
  {Butt}}{{Torres} et~al.}{2003}]{Torres2003PhR...382..303T}
{Torres} D.~F.,  {Romero} G.~E.,  {Dame} T.~M.,  {Combi} J.~A.,   {Butt} Y.~M.,
   2003, \mn@doi [\physrep] {10.1016/S0370-1573(03)00201-1}, \href
  {http://adsabs.harvard.edu/abs/2003PhR...382..303T} {382, 303}

\bibitem[\protect\citeauthoryear{{Xiao} \& {Zhu}}{{Xiao} \&
  {Zhu}}{2012}]{Xiao2012A&A...545A..86X}
{Xiao} L.,  {Zhu} M.,  2012, \mn@doi [\aap] {10.1051/0004-6361/201218938},
  \href {http://adsabs.harvard.edu/abs/2012A%26A...545A..86X} {545, A86}

\bibitem[\protect\citeauthoryear{{Xu}, {Han}, {Sun}, {Reich}, {Xiao}, {Reich}
  \& {Wielebinski}}{{Xu} et~al.}{2007}]{Xu2007A&A...470..969X}
{Xu} J.~W.,  {Han} J.~L.,  {Sun} X.~H.,  {Reich} W.,  {Xiao} L.,  {Reich} P.,
  {Wielebinski} R.,  2007, \mn@doi [\aap] {10.1051/0004-6361:20077549}, \href
  {http://adsabs.harvard.edu/abs/2007A%26A...470..969X} {470, 969}

\bibitem[\protect\citeauthoryear{{Zhao}, {Jiang}, {Gao}, {Li}  \& {Sun}}{{Zhao}
  et~al.}{2018}]{Zhao2018ApJ}
{Zhao} H.,  {Jiang} B.,  {Gao} S.,  {Li} J.,   {Sun} M.,  2018, \mn@doi [\apj]
  {10.3847/1538-4357/aaacd0}, \href
  {http://adsabs.harvard.edu/abs/2018ApJ...855...12Z} {855, 12}

\bibitem[\protect\citeauthoryear{{Zhu} \& {Tian}}{{Zhu} \&
  {Tian}}{2014}]{Zhu2014IAUS}
{Zhu} H.,  {Tian} W.,  2014, in {Ray} A.,  {McCray} R.~A.,  eds,  IAU Symposium
  Vol. 296, Supernova Environmental Impacts. pp 378--379,
  \mn@doi{10.1017/S1743921313009915}

\makeatother
\end{thebibliography}


\bsp    
\label{lastpage}
\end{document}